\documentclass[sn-nature]{sn-jnl}

\usepackage{graphicx}%
\usepackage{floatrow}
\usepackage{bm}
\usepackage{multirow}%
\usepackage{subcaption}
\usepackage{siunitx}
\usepackage{amsmath,amssymb,amsfonts}%
\usepackage{amsthm}%
\usepackage{mathrsfs}%
\usepackage[title]{appendix}%
\usepackage[percent]{overpic}
\usepackage[table]{xcolor}  
\usepackage{textcomp}%
\usepackage{manyfoot}%
\usepackage{booktabs}%
\usepackage{xcolor}
\usepackage{algorithm}%
\usepackage{algorithmicx}%
\usepackage{algpseudocode}%
\usepackage{listings}%
\usepackage{subcaption}
\usepackage{natbib}
\usepackage{multibib}
\usepackage{lineno}
\newcites{method}{ }
\newcites{main}{ }
\newcites{supp}{ }
\usepackage{setspace} 
\usepackage{lineno}
\usepackage{url}
\begin{document}

\title[Article Title]{Observation of quantum effects on radiation reaction in strong fields}


\author*[1]{\fnm{E. E.} \sur{Los}} 

\author[1, 2, 3]{\fnm{E.} \sur{Gerstmayr}}

\author[4]{\fnm{C.} \sur{Arran}}

\author[2]{\fnm{M. J. V.} \sur{Streeter}}
\author[1, 5]{\fnm{C.} \sur{Colgan}}%
\author[4, 1]{\fnm{C. C.} \sur{Cobo}}%
\author[1]{\fnm{B.} \sur{Kettle}}%
\author[6]{\fnm{T. G.} \sur{Blackburn}}%

\author[7]{\fnm{N.} \sur{Bourgeois}}%
\author[2]{\fnm{L.} \sur{Calvin}}%
\author[8]{\fnm{J.} \sur{Cardarelli}}%
\author[2]{\fnm{N.} \sur{Cavanagh}}%
\author[7]{\fnm{S.J.D.} \sur{Dann}}%

\author[9,10,11]{\fnm{A.} \sur{Di Piazza}}%

\author[8]{\fnm{R.} \sur{Fitzgarrald}}

\author[12]{\fnm{A.} \sur{Ilderton}}%

\author[11]{\fnm{C. H.} \sur{Keitel}}%

\author[6]{\fnm{M.} \sur{Marklund}}%
\author[13]{\fnm{P.} \sur{McKenna}}%

\author[4]{\fnm{C. D.} \sur{Murphy}}%

\author[1]{\fnm{Z.} \sur{Najmudin}}%

\author[7,2]{\fnm{P}. \sur{Parsons}}%

\author[7]{\fnm{P. P.} \sur{Rajeev}}
\author[7]{\fnm{D. R.} \sur{Symes}}%
\author[11]{\fnm{M.} \sur{Tamburini}}%

\author[8]{\fnm{A. G. R.} \sur{Thomas}}
\author[1, 14]{\fnm{J. C.} \sur{Wood}}%

\author[15, 16, 17]{\fnm{M.} \sur{Zepf}}%

\author[2]{\fnm{G.} \sur{Sarri}}%

\author[4]{\fnm{C. P.} \sur{Ridgers}}%

\author[1]{\fnm{S. P. D} \sur{Mangles}}%




\affil*[1]{\orgdiv{The John Adams Institute for  Accelerator Science}, \orgname{Imperial College  London}, \orgaddress{\street{Blackett Laboratory}, \city{London}, \postcode{SW72AZ}, \country{UK}}}

\affil[2]{\orgdiv{School of Mathematics and Physics}, \orgname{Queen's University Belfast}, \orgaddress{\city{Belfast}, \postcode{BT7 1NN}, \country{UK}}}

\affil[3]{\orgdiv{Stanford PULSE Institute}, \orgname{SLAC National Accelerator Laboratory}, \orgaddress{\city{Menlo Park}, \postcode{CA 94025}, \country{USA}}}

\affil[4]{\orgdiv{York Plasma Institute}, \orgname{University of York}, \orgaddress{\street{School of Physics, Engineering and Technology}, \city{York}, \postcode{YO10 5DD}, \country{UK}}}

\affil[5]{\orgdiv{Tokamak Energy Ltd}, \orgaddress{\city{Milton Park}, \postcode{OX14 4SD}, \country{UK}}}

\affil[6]{\orgdiv{Department of Physics}, \orgname{University of Gothenburg}, \orgaddress{\city{Gothenburg}, \postcode{SE-41296}, \country{Sweden}}}

\affil[7]{\orgdiv{Central Laser Facility}, \orgname{STFC Rutherford Appleton Laboratory}, \orgaddress{\city{Didcot}, \postcode{OX110QX}, \country{UK}}}

\affil[8]{\orgdiv{G\'{e}rard Mourou Center for  Ultrafast Optical Science}, \orgname{University of Michigan}, \orgaddress{\city{Ann Arbor}, \postcode{MI 48109-2099}, \country{USA}}}

\affil[9]{\orgdiv{Department of Physics and Astronomy}, \orgname{University of Rochester}, \orgaddress{\city{Rochester}, \postcode{New York 14627}, \country{USA}}}

\affil[10]{\orgdiv{Laboratory for Laser Energetics}, \orgname{University of Rochester}, \orgaddress{\city{Rochester}, \state{New York}, \postcode{14623}, \country{USA}}}

\affil[11]{\orgdiv{Max Planck Institute for Nuclear Physics}, \orgaddress{\street{Saupfercheckweg 1}, \city{Heidelberg}, \postcode{69117}, \country{Germany}}}

\affil[12]{\orgdiv{Higgs Centre}, \orgname{University of Edinburgh}, \orgaddress{\city{Edinburgh}, \postcode{EH9 3FD}, \country{UK}}}


\affil[13]{\orgdiv{Department of Physics}, \orgname{SUPA}, \orgaddress{\street{University of Strathclyde}, \city{Glasgow}, \postcode{G4 0NG}, \country{UK}}}

\affil[14]{\orgdiv{Deutsches Elektronen-Synchrotron DESY}, \orgaddress{ \city{Hamburg}, \postcode{D-22607}, \country{Germany}}}

\affil[15]{\orgdiv{Institute of Optics and Quantum Electronics}, \orgaddress{\street{Friedrich Schiller Universit\"{a}t Jena}, \city{Jena}, \country{Germany}}}

\affil[16]{\orgdiv{Helmholtz Institute Jena}, \orgaddress{\city{Jena}, \country{Germany}}}

\affil[17]{\orgdiv{GSi GmbH}, \orgaddress{\city{Darmstadt}, \country{Germany}}}

\abstract{Radiation reaction, the force experienced by an accelerated charge due to radiation emission, has long been the subject of extensive theoretical and experimental research. Experimental verification of a quantum, strong-field description of radiation reaction is fundamentally important, and has wide-ranging implications for astrophysics, laser-driven particle acceleration, next-generation particle colliders and inverse-Compton photon sources for medical and industrial applications. However, the difficulty of accessing regimes where strong field and quantum effects dominate inhibited previous efforts to observe quantum radiation reaction in charged particle dynamics with high significance.

We report the first high significance ($>5\sigma$) observation of strong-field radiation reaction on electron spectra where quantum effects are substantial. We obtain the first, quantitative, strong evidence favouring the quantum-continuous and quantum-stochastic models over the classical model; the quantum models perform comparably. The lower electron energy losses predicted by the quantum models accounts for their improved performance. Model comparison was performed using a novel Bayesian framework which has widespread utility for laser-particle collision experiments, including those utilising conventional accelerators, where some collision parameters cannot be measured directly.}

\noindent\textbf{This preprint has not undergone peer review or any post-submission improvements or corrections. The Version of Record of this article is published in Nature Communications, and is available online at \url{https://doi.org/10.1038/s41467-025-67918-8}.}

\maketitle

\section{Introduction}\label{sec2}

Quantum effects dominate charge dynamics and radiation production~\citemain{Fedotov_2023}\citemain{DiPiazza_2012} for charges accelerated by fields with strengths approaching the Schwinger field, $E_{sch}=\SI{1.3e18}{\volt\per\metre}$~\citemain{Ritus_1985}. Such fields exist in extreme astrophysical environments such as pulsar magnetospheres~\citemain{Timokhin_2010}, may be accessed by high-power laser systems~\citemain{Willingale_2023,Doria_2020,Gan_2021}, dense particle beams interacting with plasma~\citemain{Matheron_2023}, crystals~\citemain{Wistisen_2018}, and at the interaction point of next generation particle colliders~\citemain{Yakimenko_2019}. Radiation reaction affects the energy of inverse Compton scattered (ICS) photons used for various applications~\citemain{Harding_2010,Placidi_2017,Tashima_2022,Albert_2011,Geddes_2015,Gibson_2010,Telnov_2020}.
%
%
Classical radiation reaction theories do not limit the frequency of radiation emitted by accelerating charges and omit stochastic effects inherent in photon emission~\citemain{LL_1971_ClassicalTheoryFields}, thus demanding a quantum treatment. Two quantum radiation reaction models, the quantum-continuous~\citemain{Baier_1998} and quantum-stochastic~\citemain{Gonoskov_2022} models, correct the former issue, while only the quantum-stochastic model incorporates stochasticity~\citemain{Baier_1998}. Such models are of fundamental importance, providing insight into the effect of the electron self-force on its dynamics in electromagnetic fields.
In astrophysics, radiation reaction is predicted to limit electron-positron cascades which populate the magnetospheres of pulsars, magnetars and active black holes with plasma~\citemain{Timokhin_2010}\citemain{Philippov_2018} and can strongly affect reconnection in such plasmas~\citemain{Lyubarskii_1996,Uzdensky_2014}
. Radiation reaction has been proposed as a dominant factor in gamma-burst generation~\citemain{Sultana_2013} and is expected to influence the dynamics of pair-plasmas~\citemain{Zhang_2020}, including relativistic current sheets~\citemain{Jaroschek_2009}.
Strong-field quantum radiation reaction may substantially affect the interaction point at high luminosity $>\SI[]{100}{\giga\electronvolt}$ class particle colliders~\citemain{Yakimenko_2019}. Strong electromagnetic fields produced by multi-petawatt laser systems ~\citemain{Willingale_2023,Doria_2020,Gan_2021,Weber_2017,Zou_2015} will enable the exploration of compact particle acceleration~\citemain{Borghesi_2006}\citemain{Tajima_2020} and radiation generation~\citemain{Albert_2016} (e.g. via inelastic electron-photon scattering, termed Compton scattering~\citemain{Compton_1923}) in higher-power regimes. 


In strong-field environments, quantum radiation reaction is expected to dominate laser-solid target interactions~\citemain{Nakamura_2012,Zhidkov_2002}, ion-acceleration~\citemain{Capdessus_2015} and inverse Compton scattering (ICS), which has recently garnered considerable interest~\citemain{Salgado_2022,Abramowicz_2021}.

Understanding the effect of radiation reaction on ICS photon spectra is crucial for diverse applications including industrial, defence, archaeological~\citemain{Harding_2010,Placidi_2017}, and medical~\citemain{Tashima_2022} imaging, nuclear physics~\citemain{Albert_2011,Geddes_2015,Gibson_2010}, and proposed schemes for gamma-gamma colliders~\citemain{Telnov_2020}.

The impact of radiation reaction on particle dynamics is characterised by the dimensionless intensity parameter, $a_0=\frac{E_Le}{\omega_L m_ec}$, and the electron quantum parameter, $\eta=E_{RF}/E_{sch}$, where $e$ and $m_e$ are the electron charge and mass respectively, $c$ is the speed of light in vacuum, $E_L$, $E_{RF}$ are the external electric field (laser) strengths in the laboratory and electron rest frames, respectively, and $\omega_L$ is the electric field frequency in the lab frame. When $a_0 \gtrsim 1$, both relativistic and multi-(laser) photon effects become important. The regime of strong classical radiation reaction is characterised by $\alpha a_0\eta\simeq 1$ and $\eta \ll 1$, where $\alpha$ is the fine structure constant~\citemain{DiPiazza_2012}\citemain{Blackburn_2020_b}. Quantum effects dominate when $\alpha a_0\simeq 1$ and $\eta\gtrsim1$~\citemain{DiPiazza_2012}\citemain{Blackburn_2020_b}.

In regimes dominated by classical radiation reaction, an electron emits many photons which each remove a small fraction of its energy~\citemain{Blackburn_2014}; radiation emission is treated as continuous and its impact on electron motion is well-described by the classical Landau-Lifshitz equation~\citemain{LL_1971_ClassicalTheoryFields}. In the strong-field quantum regime, interactions with the laser field must be treated non-perturbatively and are absorbed into electron basis states by quantizing the Dirac field in the presence of the laser field (Furry picture). Photon emission is described perturbatively with respect to these states~\citemain{Fedotov_2023}\citemain{DiPiazza_2012}, and becomes stochastic, with single emissions
removing significant fractions of the electron energy~\citemain{DiPiazza_2012}. In this work, the ``quantum-stochastic'' model of radiation reaction employs the locally constant field approximation (LCFA), which assumes emission events are point-like, depending only on local electric and magnetic fields, which are assumed to be constant over the timescale of photon emission~\citemain{Ritus_1985}. Between emission events, electron motion is assumed to be classical, which is a good approximation in the ultra relativistic regime~\citemain{Kirk_2009}.


The quantum-continuous model, known as the semi-classical model in the literature, aims to incorporate quantum physics in a classical framework, treating radiation emission as continuous~\citemain{Ridgers_2017}, but capturing the same rate of change of average electron momentum as the quantum-stochastic model via the inclusion of a correction term, the Gaunt factor~\citemain{Baier_1998}.



To date, six studies~\citemain{Wistisen_2018}\citemain{Poder_2018,Cole_2018,Mirzaie_2024,Matheron_2024} have aimed to measure radiation reaction in strong fields ($a_0\gg1$). Of these studies, only one~\citemain{Cole_2018} conducted a quantitative comparison of different radiation reaction models, which was limited to $1\sigma$ significance. 

Using an all-optical setup, we accessed $a_0\approx10$ and $\eta\leq0.09$, for which strong-field non-perturbative effects dominate and quantum effects are influential, in contrast to early experiments at LINACs~\citemain{Bula_1996,Burke_1997,Bamber_1999} and storage rings~\citemain{Sands_1960}, for which $a_0<1$. Previous experiments using crystals~\citemain{Wistisen_2018}\citemain{Wistisen_2019,Nielsen_2020}, at CORELS~\citemain{Mirzaie_2024} or at ELI-NP and APPOLLON~\citemain{Matheron_2024} did not report an observation of radiation reaction on lepton spectra. 

This work is the first to exceed the $5\sigma$ significance threshold required for a definitive observation of radiation reaction, and the first to present strong, quantitative evidence favouring quantum models over a classical model. This represents a substantial improvement on previous all-optical experiments which reported $\leq 3 \sigma$ evidence of radiation reaction and performed model selection with $<1\sigma$ significance due to data scarcity and large uncertainties~\citemain{Cole_2018,Poder_2018}. The higher significance of our findings is largely due to the greater number of successful collisions reported in this work ($>600$) compared to previous all-optical experiments ($<10$). This substantial increase in the number of successful collisions was enabled by the implementation of automated timing and pointing stabilisation for both lasers.

Notably, our findings clarify outstanding questions raised by previous experiments~\citemain{Cole_2018,Poder_2018} regarding the relative validity of different radiation reaction models in the classical-quantum regime.

\begin{figure}[ht!]%
 \centering
 \includegraphics[width=0.99\textwidth, trim={0.65cm 0.5cm 0.0cm 1.0cm}, clip]{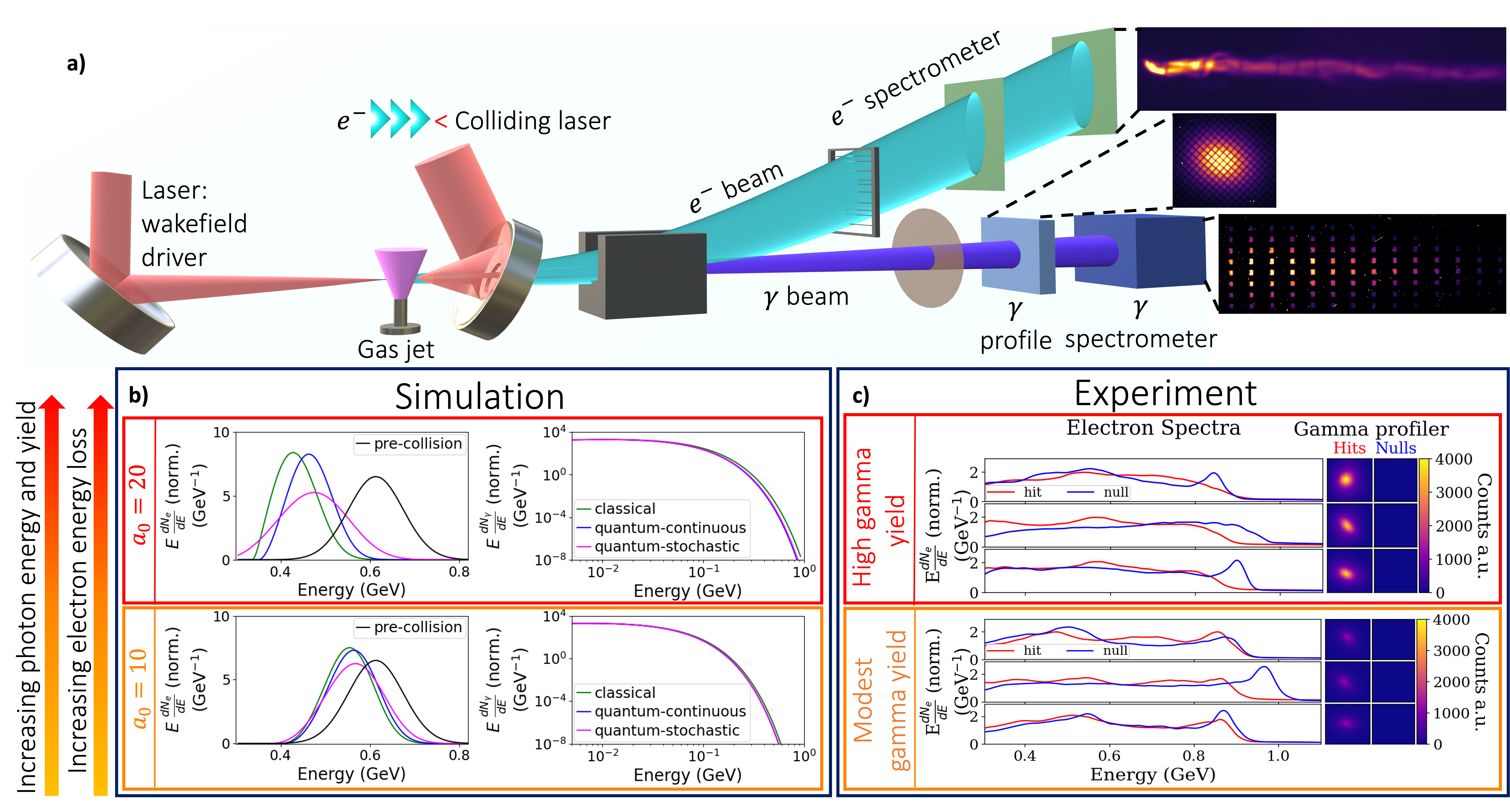}
 \caption{\textbf{Experimental set-up, qualitative comparisons of measured hits and nulls and simulated radiation reaction models.} a) Experimental setup: one laser pulse, focused into a gas jet, drove a wakefield accelerator. A second, tightly focused, counter-propagating laser pulse collided with the electron beam which emitted gamma photons. The electron spectrometer consisted of a dipole magnet which dispersed the electron beam through a wire array onto two LANEX scintillating screens (green). A Caesium Iodide (CsI) profile screen and stack characterised the transverse profile and spectrum of the emitted gamma radiation, respectively. b) Simulated post-collision electron spectra (normalised by integration) and photon spectra illustrating the classical, quantum stochastic an quantum-continuous model predictions for $a_0=10$ (bottom) and $a_0=20$ (top). The electron beam and laser pulse collided \SI[]{40}{\femto\second} after focus. The transverse and longitudinal laser intensity profiles were gaussian, with respective full-width half-maxima (FWHM) of \SI[]{2.47}{\micro\metre} and \SI[]{30}{\femto\second}. c) Measured electron spectra for hits with high gamma profile yields are shown above those measured for and moderate yields, together with corresponding gamma profile signals. Nulls have been randomly selected.\label{fig:exp_setup}}
 \end{figure}

The experimental setup is illustrated in figure~\ref{fig:exp_setup}a). Electron beams with mean energy $\approx\qty[parse-numbers = false]{609  \pm 2(stat) (12)(syst)}{\mega\electronvolt}$ and shot-to-shot standard deviation \SI[separate-uncertainty]{40(1)}{\mega\electronvolt} (standard errors were calculated using bootstrapping due to non-normal data distributions), generated using laser-driven wakefield acceleration~\citemain{Tajima_2020}, collided with a tightly focused, counter-propagating laser pulse with $I=\SI[separate-uncertainty]{1.0(2)e21}{\watt\per\centi\metre\squared}$, $\lambda_L=2\pi c/\omega_L=0.8\;\text{$\mu$m}$, $a_0= 21.4\pm1.8$ and $\langle\eta\rangle\leq0.13\pm0.02$ at the laser focus, sufficient to probe the strong-field quantum regime. For more details, see experimental methods~\ref{sec:exp_methods}.

Figure~\ref{fig:exp_setup}b) compares predictions for the post-collision electron and photon spectra for the classical, quantum-continuous and quantum-stochastic models. All models predict net electron energy losses (evidenced by lower-energy post-collision electron spectra), which, together with photon yield, scale with increasing $a_0$. These model-independent indicators of radiation reaction are used in the frequentist analysis in section~\ref{sec:results}.

Quantum models prohibit the emission of photons with energies exceeding the electron energy, thus predicting lower energy losses than the classical model, evidenced by the mean post-collision electron energies and photon yields above \SI{100}{\mega\electronvolt} in figure~\ref{fig:exp_setup}b). Unlike the classical and quantum-continuous models, the quantum-stochastic model predicts spectral broadening, arising from probabilistic photon emission~\citemain{Neitz_2013,Ridgers_2017,Niel_2018}, indicated by the relative widths of post-collision electron spectra in figure~\ref{fig:exp_setup}b).

The model-independent trends in figure~\ref{fig:exp_setup}b) appear qualitatively in experimental data in figure~\ref{fig:exp_setup}c); this is purely illustrative and does not represent the frequentist analysis. In figure~\ref{fig:exp_setup}c), electron spectra for hits with moderate and high photon yields exhibit less pronounced high-energy peaks than randomly selected nulls, consistent with fewer high-energy electrons.  As photon yield increases, the proportion of charge at high energies decreases. Some nulls have lower energies than hits, illustrating the shot-to-shot variability in electron energy.


\begin{figure}[ht!]
\centering
{\begin{overpic}[width=0.99\textwidth, trim={0.5cm, 0.2cm, 0.2cm, 0.5cm},clip]{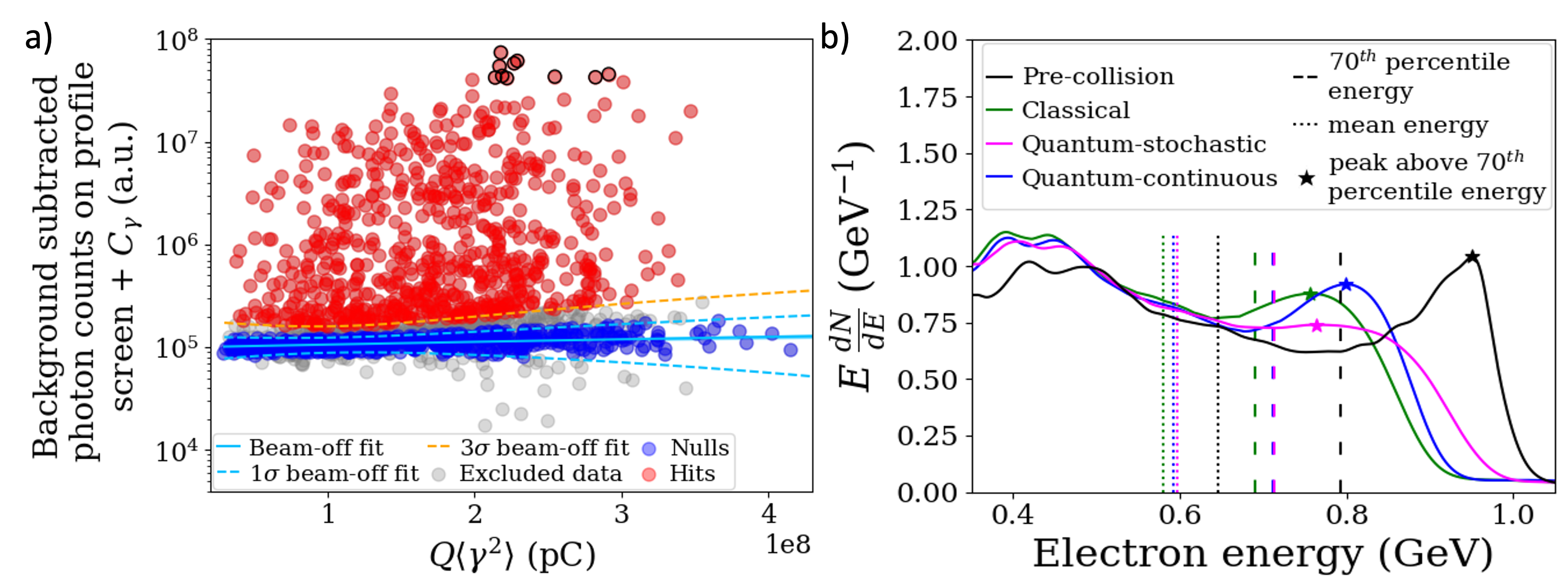}
\end{overpic}}
\caption{\textbf{Shot selection and summary statistics}
a) The shot selection procedure is illustrated. Background-subtracted total counts measured by the gamma profile diagnostic are shown as a function of $Q\langle\gamma^2\rangle$ for all shots, where $Q$ and $\gamma$ denote electron beam total charge and Lorentz factor, respectively. A constant, $C_\gamma=1\times10^{5}$ has been added to the normalised total counts for all shots to allow the data to be shown on a logarithmic scale. Nulls (blue, 608 shots) consist of combined misses and beam-off shots. The latter lie within $1\sigma$ (cyan, dashed) of the scaling of background gamma yield with $Q\langle\gamma^2\rangle$ (cyan, continuous). The small fraction of nulls which lie above this threshold are beam-off shots.
Hits (red, 687 shots) lie $3\sigma$ (orange, dashed) above the background scaling. The grey points cannot be categorised as hits or nulls and thus are excluded from the analysis. The shots analysed using the Bayesian framework are encircled (black). b) Simulated post-collision electron spectra, normalised by integration, predicted by different radiation reaction models for a collision between a electron beam (pre-collision spectrum shown) with a gaussian temporal profile with full-width half-maxima (FWHM) \SI[]{141}{\femto\second} and a laser pulse with $a_0 = 14$ and gaussian transverse and longitudinal intensity profiles with FWHM \SI[]{2.47}{\micro\metre} and \SI[]{45}{\femto\second}, respectively. The collision was offset temporally from the laser focus by \SI[]{60}{\femto\second}. The mean energy, $\langle E \rangle$ and peak height above the $70^{th}$ percentile electron energy, $P_{70}$, which indicates the prominence of the high energy peak in the spectrum, are shown.
\label{fig:gp_shot_sel}}
\end{figure}

\section{Results}\label{sec:results}

In lieu of measured pre-collision electron spectra, electron beams measured for misses and beam-off shots (nulls) are compared to those measured for successful collisions (hits). Hit and null identification is illustrated in figure~\ref{fig:gp_shot_sel}a) and discussed in section~\ref{subsec:shot_selection}. The profile screen background subtraction is detailed by Los~\citemain{Los_2023}. 

Figure~\ref{fig:gp_shot_sel}b) illustrates two model-independent signatures of radiation reaction, identified by extensive simulation work provided in figure~\ref{fig:peakiness_demo} in the Extended Material~\ref{sup:shot_selection_fig}, and as shown by Los et al.~\citemain{Los_2024}. These are used to quantitatively examine whether differences between hits and nulls exceeded shot-to-shot variation in electron spectra, and whether any such differences are consistent with radiation reaction. Compared to the pre-collision spectrum, simulated post-collision spectra have lower mean energies, $\langle E \rangle$, and less pronounced peaks above the $70^{th}$ percentile energy, $P_{70}$, calculated using electron spectra normalised by integration. The $70^{th}$ percentile energy is the energy under which $70\%$ of the electron beam charge lies. Changes in $P_{70}$ reflect the re-distribution of charge due to spectral broadening. Spectral broadening may reflect stochasticity, or spatial and temporal misalignments between the laser pulse and electrons, causing equally energetic electrons to experience different laser intensities and hence energy losses. 

\begin{figure}[ht!]%
\centering
{\begin{overpic}[width=1.0\textwidth, trim={0.2cm 0.0cm 0.0cm 0.0cm}, clip]{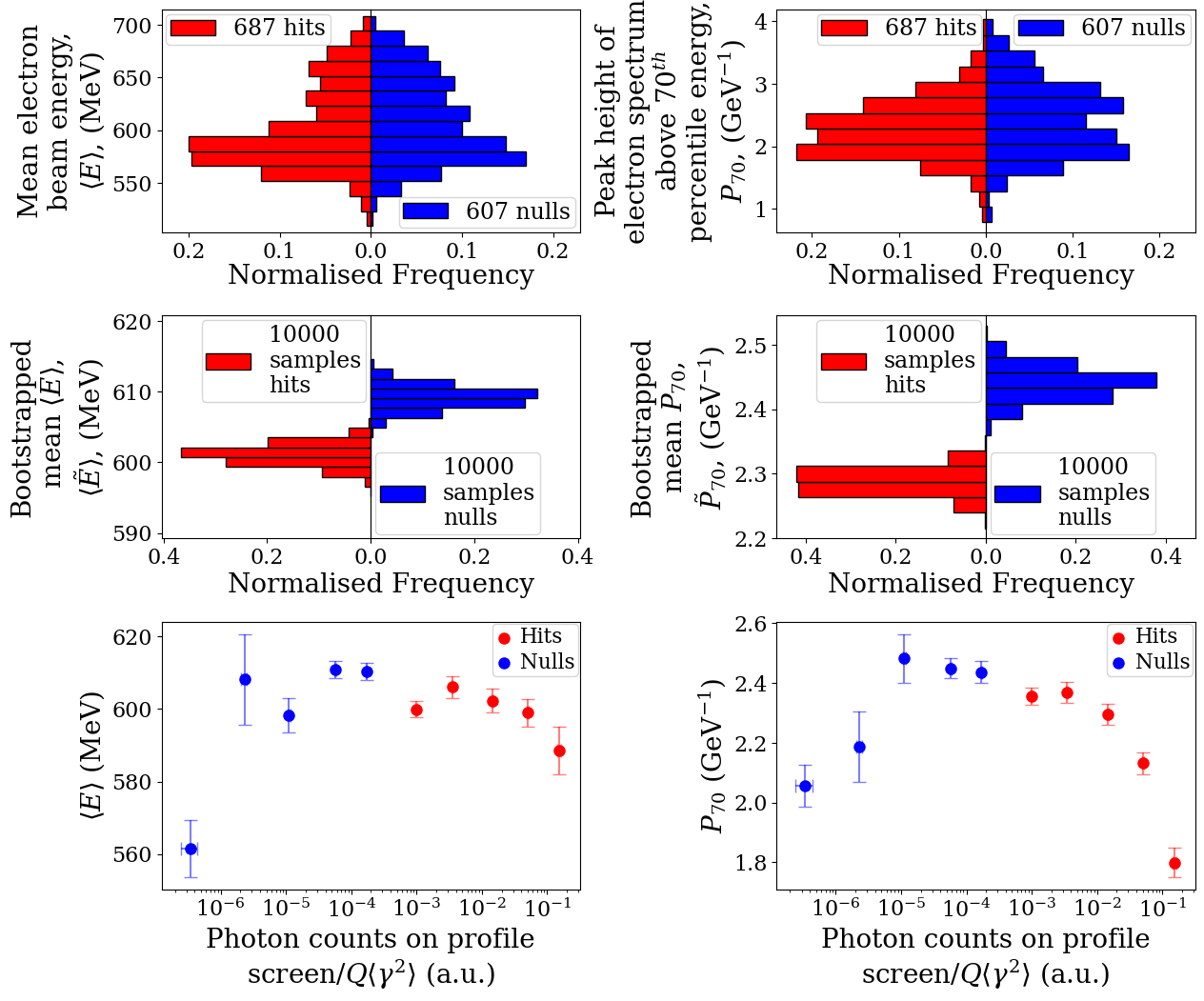}
\put(-1,84){a)}
\put(49,84){b)}
\put(-1,57){c)}
\put(49,57){d)}
\put(-1,32){e)}
\put(49,32){f)}
\end{overpic}}
\caption{\textbf{Model-independent analysis of electron energy loss and photon yield.} 
Distributions of a) $\langle E \rangle$ and b) $P_{70}$, for measured hits (red) and nulls (blue). Hit and null distributions have been normalised to the total number of shots in each.
Hit and null distributions of c) mean $\langle E \rangle$, denoted $\langle \tilde{E} \rangle$, and d) mean $P_{70}$, denoted $\tilde{P}_{70}$, obtained by bootstrapping hit and null distributions in a) and b), respectively. 
Bottom: e) Mean $\langle E \rangle$ for the 687 hits and 607 nulls analysed, binned logarithmically by gamma profile yield normalised to $Q\langle\gamma^2\rangle$; different bins contain different numbers of shots. f) Similar to e), for $P_{70}$. 
\label{fig:eloss_hists}}
\end{figure}

In figures~\ref{fig:eloss_hists}a) and b), there are fewer high-energy electron spectra, and fewer strongly-peaked spectra for hits compared to nulls, meaning fewer high-energy and more low-energy electron beams in the former population compared to the latter. Two-sample Kolmogorov-Smirnov (KS) tests (selected for applicability to arbitrary distributions) confirm this, indicating the null hypothesis that hits and nulls originated from the same distribution can be rejected at the $5\sigma$ (p-value $=5.3\times10^{-9}$) and $4\sigma$ (p-value $=2.7\times10^{-5}$) levels for $P_{70}$ and $\langle E \rangle$, respectively.

The distribution of mean values, $\langle \tilde{E} \rangle$, calculated for the distribution of $\langle E\rangle$ in figure~\ref{fig:eloss_hists}a) using bootstrapping (see section~\ref{Sec:bootstrapping}) is shown in figure~\ref{fig:eloss_hists}c). The distribution of mean values for $P_{70}$, denoted $\tilde{P}_{70}$, is shown in figure~\ref{fig:eloss_hists}b). The hit distribution means for $\langle E\rangle$ and $\tilde{P}_{70}$ lie $3\sigma$ (p-value $=2.0\times10^{-4}$) and $5\sigma$ (p-value $=3.3\times10^{-7}$) below the corresponding means for nulls, respectively. The close agreement between the significance values obtained using KS tests and by bootstrapping attests to their reliability. The highly significant observation of lower mean electron energies and lower peak heights for hits compared to nulls is consistent with electron energy loss and spectral broadening, and constitutes strong evidence of radiation reaction.


As illustrated in figure~\ref{fig:gp_shot_sel}a) photon yields for all background sources (betatron radiation, bremsstrahlung), and thus for nulls, scale positively with $Q\langle\gamma^2\rangle$, where $Q$ and $\langle\gamma^2\rangle$ denote the total charge and expected value of $\gamma^2$ for the electron beam. The photon counts in figures~\ref{fig:eloss_hists}e) and f) are normalised by $Q\langle\gamma^2\rangle$ to remove this background scaling to first-order. After this correction is applied, a residual positive correlation remains for the nulls.

For hits in figures~\ref{fig:eloss_hists}e) and f), $P_{70}$ and $\langle E \rangle$ decrease with increasing normalised photon yield. For the brightest hits, $P_{70}$ and $\langle E \rangle$ lie $15\sigma$ and $7\sigma$ below the null means, respectively. Thus, the electron beam has lower energy for successful collisions compared to unsuccessful collisions, consistent with radiation reaction.

By contrast to the positive scaling observed for nulls, hits exhibit a negative scaling with gamma yield, consistent with energy loss and hence radiation reaction. The correlation coefficients for hits and nulls differ significantly; by $40\sigma$ and $34\sigma$ for $P_{70}$ and $\langle E \rangle$, respectively (see figure~\ref{fig:corr_coeffs}, supplementary note~\ref{supp_freq_anal}). The negative correlations observed for hits demonstrate that different physical mechanisms dominate radiation production compared to nulls and confirm that energy losses ``switch-on'' for successful collisions. 

The highly significant observation of electron energy loss which increases with photon yield constitutes the first observation of radiation reaction on particles.

All-optical experiments offer advantages such as natural timing and alignment of the electron beam and the colliding laser, but also present challenges. Parameters such as laser intensity and the relative size, duration and spatio-temporal overlap of the electron beam and the colliding laser pulse strongly affect post-collision electron and photon spectra, but are not measured on-shot and vary substantially between shots. This precludes the straightforward model comparison illustrated in figure~\ref{fig:exp_setup}b), and necessitates an analysis which accounts for uncertainties due to unknown pre-collision electron spectra and unknown collision parameters when comparing radiation reaction models. This was achieved using Bayesian inference.

If a model, $M$, depends on parameters with unknown values, Bayesian inference estimates these parameters by constructing the probability that the model is accurate given the data observed, $D$, called the posterior probability, $P(M|D)$ (this notation integrates over all parameter values). The most likely parameter values are those which optimise the posterior probability.


The posterior probability is calculated using Bayes' theorem, given in equation~\ref{eqn:Bayes_theorem} ~\citemain{Heard_2021},
\begin{equation}
\label{eqn:Bayes_theorem}
P(M|D)=\frac{P(D|M)P(M)}{P(D)}
\end{equation}
where the likelihood, $P(D|M)$, is the probability of observing the data given the model, the prior probability, $P(M)$, represents prior beliefs about the model before observing data and $P(D)$ is the (constant) probability of observing the data. In the absence of an analytic expression for the posterior, the posterior distribution is computed algorithmically by iteratively sampling model parameters, calculating the corresponding likelihood and hence updating the posterior until convergence is reached. For efficiency, high-probability regions of the parameter space are preferentially sampled. 




As pre-collision electron spectra cannot be measured for successful collisions, distributions of pre-collision electron spectra were predicted for these shots using a neural network~\citemain{Streeter_2023}, given the measured laser energy, plasma density and longitudinal profile of plasma re-combination light. The distribution variance reflects the prediction uncertainty, which, together with experimental uncertainties, is accounted for in the Bayesian analysis.

Collision parameters which could not be measured were inferred using Bayesian inference~\citemain{Los_2024}. To avoid over-fitting and excessive computational run times~\cite{Los_2024}, only a subset of collision parameters was inferred and all other parameters were fixed. We chose to infer $\tau_e$, $a_0$ and the longitudinal displacement of the collision from the laser focus, $Z_d$ as they have the highest expected impact on the post-collision electron and photon spectra and exhibit degeneracies with parameters which were not inferred. Degeneracies allow different combinations of collision parameters to produce the same collision distributions of $\eta$ and $a_0$, denoted by $\tilde{\eta}$ and $\tilde{a}_0$, and hence the same observables, as discussed in section~\ref{sec:BI_excl_param_summary} and by Los et al.~\citemain{Los_2024}. Hence, the inference procedure returns ``effective'' values for $\tau_e$, $a_0$ and $Z_d$; so called as they reflect the distributions of $\tilde{\eta}$ and $\tilde{a}_0$ that reproduce the observables, rather than accurately representing electron beam and laser properties. 

Transverse misalignments between the electron beam and the laser pulse, $r_d$, were assumed to be $0$. To maximise the probability that this condition was met, only the ten shots with the highest highest gamma yields normalised by $Q\langle\gamma^2\rangle$ were analysed. Constraints in computational resources limited the number of shots analysed; each inference required $\approx19200$CPU hours, 60GB per CPU. 

Three inferences were performed per shot; one for each model of radiation reaction. During each inference, different sets of collision parameters were combined with the predicted pre-collision electron spectrum to re-construct different collisions and corresponding post-collision electron and photon spectra for the relevant radiation reaction model. 

Although the ``true'' parameters are unknown, the relative validity of two models can be compared using the ratio of model evidences, or Bayes factor (see equation~\ref{eqn:bic}, section~\ref{sec:ana_methods}). The model evidence is obtained by integrating the likelihood weighted by the parameter priors over all parameter space. Bayes factors provide a more robust metric of model performance compared to a frequentist approach (e.g. a least-squares fit), which only compares the validity of a model to a null hypothesis for the ``best-fit'' parameters.

The Bayesian analysis was tested on realistic electron spectra for various simulated collisions with differing collision parameters. For each test, the Bayesian analysis consistently favoured the correct model and inferred the first moments of the collision distributions of $\tilde{\eta}$ and $\tilde{a}_0$, respectively denoted by $\langle\tilde{\eta}\rangle$ and $\langle \tilde{a}_0\rangle$, within $1\sigma$ of their simulation values. Model differentiation was only accurate for $r_d\leq1.6w_0$, where $w_0=\SI{2.5}{\micro\metre}$ was the laser waist at focus~\citemain{Los_2024}.

\begin{figure}[!ht]
    \centering
    \begin{subfigure}[t]{1.0\textwidth}
    	{\begin{overpic}[width=1.0\linewidth, trim={0.25cm, 0.0cm, 0.19cm, 0.23cm},clip]{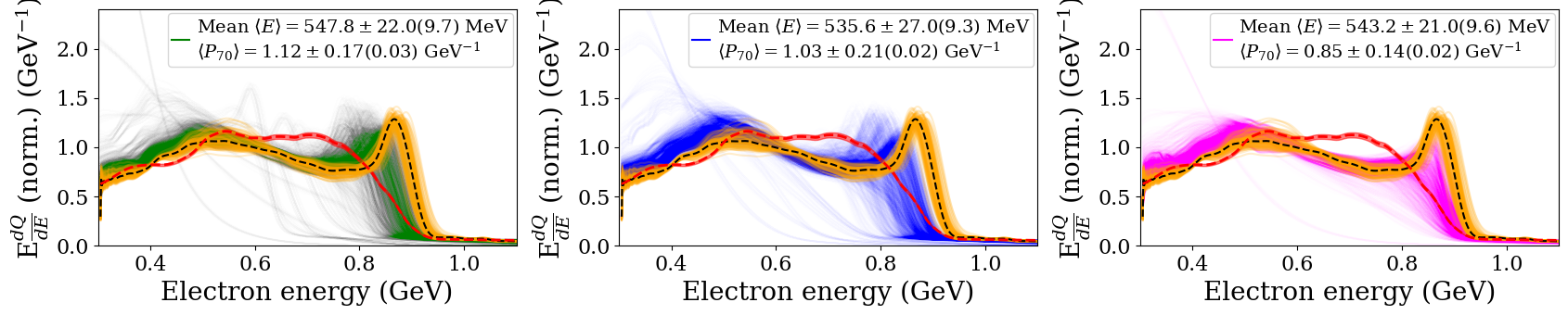}
	\put(-2.2,18){a)}
	\put(14,20.5){Classical}
    \put(40.5,20.5){Quantum-continuous}
	\put(74.5, 20.5){Quantum-stochastic}
	\end{overpic}}
    \end{subfigure}
    \vspace{0.0cm}
    \begin{subfigure}[t]{1.0\textwidth}
    	{\begin{overpic}[width=1.0\linewidth, trim={0.2cm, 0.2cm, 0.16cm, 0.22cm},clip]{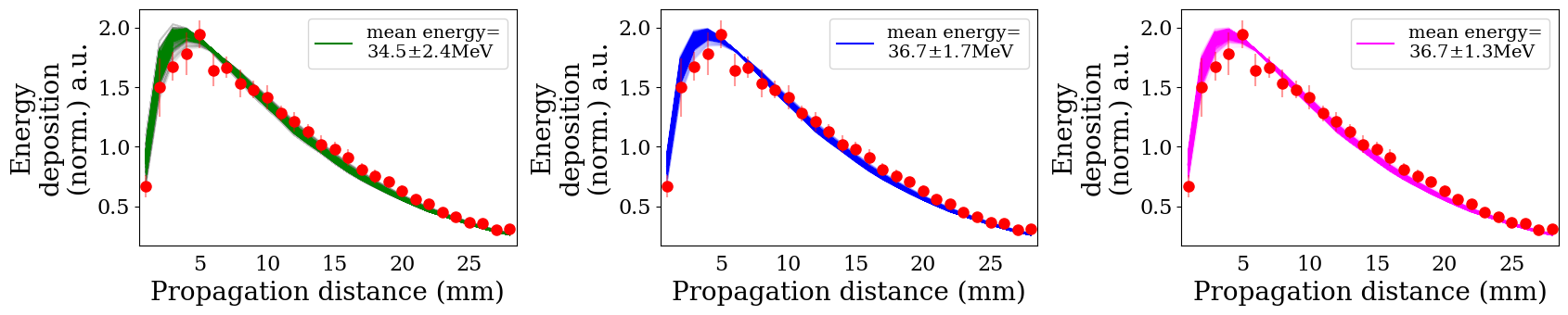}
    \put(-2.5,19.0){b)}
	\end{overpic}}
    \end{subfigure}
    \caption{\textbf{Bayesian inference results for the highest gamma-yield shot normalised to $Q\langle\gamma^2\rangle$ (shot 2).} Measured data (red) and predictions for the classical (green), quantum-continuous (blue), quantum-stochastic (magenta) models, which inferred $\langle \tilde{a}_{0}\rangle=6.2\pm1.0$ and $\sigma_{a_0}=1.2\pm0.3$, $\langle \tilde{a}_{0}\rangle=6.8\pm0.9$ and $\sigma_{a_0}=0.4\pm0.1$ and $\langle \tilde{a}_{0}\rangle=6.7\pm0.9$ and $\sigma_{a_0}=0.4\pm0.1$, respectively.
    a) Measured and inferred post-collision electron spectra. For the former, 
    $\langle E \rangle =\qty[parse-numbers = false]{(564.1\pm0.0(10.3))}{\mega\electronvolt}$
    $P_{70}=\qty[parse-numbers = false]{0.83 \pm 0.00 (0.01)}{\per\giga\electronvolt}$. The distribution of pre-collision electron spectra predicted by the neural network (orange), for which $\langle E\rangle=\qty[parse-numbers = false]{(574.1 \pm 3.9 (10.7))}{\mega\electronvolt}$, $P_{70}=\qty[parse-numbers = false]{1.29 \pm 0.05 (0.02)}{\per\giga\electronvolt}$, and its median (black).
    b) Measured and inferred photon energy deposition in each scintillation crystal as a function of propagation distance in the CsI photon spectrometer. The mean photon energy measured was \SI[separate-uncertainty]{63.3(58)}{\mega\electronvolt}.
    \label{fig:post-coll_espec_no_chirp_1}}
\end{figure}
    
\begin{figure}[!ht]
    \centering
   {\begin{overpic}[width=0.8\linewidth, trim={0.25cm, 0.2cm, 0.2cm, 0.2cm},clip]{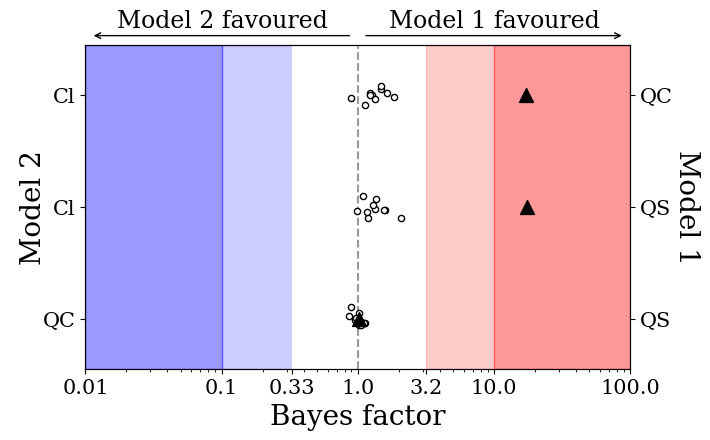}
   \end{overpic}}
    \caption{\textbf{Bayesian comparison of radiation reaction models.} 
    Bayes factors for individual shots (circles) and combined over ten shots (triangles) are shown. Weak (white), substantial (light shading) and strong (dark shading) evidence favouring model 1 (red) or model 2 (blue) are categorised according to the half-log scale convention outlined by Kass and Raftery~\protect\citemain{Kass_1995}. The dashed grey line indicates equal performance of compared models.\label{fig:Bayes_factors}}
\end{figure}

Measured and inferred post-collision electron and photon spectra are shown in figure~\ref{fig:post-coll_espec_no_chirp_1} for the highest normalised gamma yield shot. Additional results are provided in supplementary note~\ref{sup:extra_BI_results}. In figure~\ref{fig:post-coll_espec_no_chirp_1}a), the $\langle E\rangle$ and $P_{70}$ measured post-collision lie $3\sigma$ and $11\sigma$ below the neural net predictions, respectively; note systematic uncertainties cancel as the correlation coefficient is 1. Only the quantum-stochastic model infers both $\langle E\rangle$ and $P_{70}$ within $1\sigma$ of the measured spectrum, indicating marginally higher performance. Both quantum models predict more accurate photon spectra with lower uncertainties than the classical model, as indicated by the mean photon energies in figure~\ref{fig:post-coll_espec_no_chirp_1}b), typically the inferred photon energy deposition lies within $1\sigma$ of the measured energy deposition, confirming the goodness of fit for this diagnostic. The requirement to reproduce both electron and photon spectra using only three fitting parameters places a stronger constraint than an analysis considering a single diagnostic, which risks overfitting. This accounts for the discrepancy between measured and inferred mean photon energies, and highlights the importance of considering multiple diagnostics simultaneously.

In figure~\ref{fig:Bayes_factors}, individual shots yield weak (Bayes factor $<3.2$~\citemain{Kass_1995}), but consistent evidence favouring the quantum models over the classical model, increasing the credibility of the former. The Bayes factors for the quantum-continuous and quantum-stochastic models lie consistently around 1, signifying their comparable performance. 
As the inferred parameters for each shot are independent, model evidences can be combined by multiplication. The combined Bayes factors over 10 shots indicate strong evidence favouring the quantum-stochastic and quantum-continuous models over the classical model, but insufficient evidence to distinguish between the quantum models.

The quantum models better replicate both photon and electron spectra simultaneously compared to classical predictions as they predict lower energy losses. The electron spectra largely determine the posterior location, while photon spectra predominately affect the posterior shape as their likelihood distributions are far narrower; inferred photon spectra have fewer degrees of freedom and thus are less accurate than inferred electron spectra, but their fractional uncertainties are not substantially greater. Thus, both diagnostics play a critical role in constraining the posterior. As electron spectra contribute similarly to the posterior probability across all models, as indicated by their inferred $\langle E\rangle$ and $P_{70}$, relative model performance is determined by the photon spectrum. Quantum models yield higher posterior probabilities than the classical model as they infer photon spectra more accurately and with lower uncertainty. The novel ability to analyse photon and electron spectra within one self-consistent framework is critical for model comparison, and a key strength of this Bayesian approach.


The quantum-stochastic model inferred $0.05\leq\langle\tilde{\eta}\rangle\leq0.1$ and $7\leq\langle \tilde{a}_0\rangle\leq13$ across ten shots. For the inferred $\langle\tilde{a}_0\rangle\geq7$, the transverse offset was $r_d\leq0.64w_0$, well within the range of $r_d$ for which model selection is accurate.

Analytic classical and quantum-stochastic models~\citemain{Blackburn_2024_2,Vranic_2016} and a numerical quantum-continuous model provide an independent corroboration of the Bayesian analysis. A gaussian is fitted to the high-energy peak in the pre-collision electron spectrum in figure~\ref{fig:post-coll_espec_no_chirp_1}a). Using the $\langle \tilde{a}_0\rangle$ inferred by the Bayesian analysis, the corresponding post-collision spectrum is predicted analytically for each model. The mean energy, $\langle E_p \rangle$, and standard deviation, $\sigma_p$, of the pre-collision peak are $\langle E_p \rangle=\SI[]{864}{\mega\electronvolt}$ and $\sigma_p=\SI[]{32}{\mega\electronvolt}$ $(4\%)$. The post-collision peak predicted by the classical model has $\langle E_p \rangle=\SI[]{807}{\mega\electronvolt}$ and $\sigma_p=\SI[]{19}{\mega\electronvolt}$ $(2\%)$, while quantum-continuous and quantum-stochastic models
predict peaks with $\langle E_p \rangle=\SI[]{803}{\mega\electronvolt}$ and $\sigma_p=\SI[]{34}{\mega\electronvolt}$ $(4\%)$, and $\langle E_p \rangle=\SI[]{814}{\mega\electronvolt}$ and $\sigma_p=\SI[]{58}{\mega\electronvolt}$ $(16\%)$, respectively, consistent with Arran et al.~\citemain{Arran_2019} and Yoffe et al.~\citemain{Yoffe_2015}. The predicted post-collision spectra are consistent with the measured data, validating the Bayesian analysis. The post-collision spectral widths predicted analytically appear to indicate the feasibility of model selection between all three models for these collision conditions. However, as large variations in laser intensity during the collision also cause spectral broadening, more precise knowledge of collision conditions is needed to differentiate between quantum models.

The onset of quantum effects and reduced accuracy of the Landau-Lifshiftz model observed for $0.05\leq\langle\tilde{\eta}\rangle\leq0.1$ and $7\leq\langle \tilde{a}_0\rangle\leq13$ motivate the use of quantum-corrected models in this regime. Quantum-corrected radiation reaction models predict lower energy losses for charged particles in strong fields compared to classical models, with wide-ranging consequences. Quantum-corrected ICS photon spectra have fewer high energy photons; this has implications for ICS-based photon sources used for imaging and nuclear physics. In astrophysical environments, a quantum-corrected synchrotron emission reduces emission and cooling rates of pulsars and galactic nuclei jets and affects synchrotron emission-based estimates of magnetic field strengths. Quantum-corrected radiation models indicate higher centre-of-mass energies in particle accelerators, while for laser-solid target ion acceleration, such models predict stronger sheath fields and hence higher ion energies.


In conclusion, we report the first highly significant observation of radiation reaction on electron and photon spectra and present quantitative, strong evidence to favour two quantum models over a classical model for the first time, for $0.05\leq\langle\tilde{\eta}\rangle\leq0.1$ and $7\leq\langle \tilde{a}_0\rangle\leq13$, due to the lower energy losses the former predict. While spectral broadening was observed, insufficient knowledge of collision parameters and large uncertainties on predicted pre-collision electron spectra inhibited our ability to determine whether this arose due to stochasticity. Model differentiability would improve with stable, mono-energetic electron beams and strongly constraining priors motivated by collision parameter measurements. Lower variation in transverse alignment or in collisions with an expanded, higher-power laser would boost statistics at higher $a_0$, facilitating investigations of quantum phenomena over a greater range of $\eta$ and providing new insight into the validity of the quantum-continuous and quantum-stochastic models in these regimes.

\clearpage

\bibliographystylemain{unsrt}
\bibliographymain{main}

\clearpage

\section{Methods}\label{sec:methods}
\subsection{Experimental Methods}\label{sec:exp_methods}


The experiment was conducted using the dual-beam Gemini laser at the Central Laser Facility, Rutherford Appleton Laboratory, UK. An \textit{f}/40 off-axis parabola was used to focus one arm of the linearly polarised, two-beam system to a transverse full-width half-maximum (FWHM) of the focal spot intensity of \SI[separate-uncertainty]{35(3)}{\micro\metre}$\times$\SI[separate-uncertainty]{40.6(12)}{\micro\metre}. An off-shot Grenouille measurement of the FWHM duration of the laser intensity yielded \SI[separate-uncertainty]{59.5(25)}{\femto\second}. The laser delivered \SI[separate-uncertainty]{6.8(06)}{\joule} to target, corresponding to an $a_0=1.0\pm0.15$ (standard deviation given).

The laser-wakefield drive beam was focused into a \SI[]{15}{\milli\metre} supersonic gas jet with a trapezoidal density profile and \SI[]{5}{\milli\metre} ramps. An average electron density of $\approx\SI[separate-uncertainty]{1.1(2)e18}{\per\centi\metre\cubed}$ was attained at the peak of the trapezoidal profile with He gas, doped with $1\%$ $N_2$ to induce ionisation injection~\citemethod{McGuffey_2010}. 
The electron beam, and photons produced by inverse Compton scattering (ICS), propagated through the hole in the \textit{f}/2 parabola, which had an acceptance angle of \SI[]{42}{\milli\radian}. A $\int B(x)dx$=\SI[]{0.4}{\tesla\metre} dipole magnet was used to disperse the electron beam through a wire array onto two sequential LANEX screens which were imaged by two cooled 16-bit cameras. The electron spectrum was subsequently retrieved with the aid of a tracking algorithm which computed the trajectories of electrons through the magnetic field. The wire array and two screens allowed the degeneracy between the electron beam energy and pointing into the magnet to be de-convolved~\citemethod{Clayton_2010}\citemethod{Wang_2013}\citemethod{Soloviev_2011}. The systematic uncertainty in the retrieved electron energy due the uncertainties in the relative positions of the magnet, lanex screens and gas jet was
\begin{equation}
\label{eqn:electron_syst_energy}
\zeta_{e}[\SI[]{}{\mega\electronvolt}]=C_eE[\SI[]{}{\mega\electronvolt}]^2,
\end{equation}
where $E$ denotes electron energy and $C_e=\SI{32.45e-6}{\per\mega\electronvolt}$.

The wakefield accelerator produced \SI[separate-uncertainty]{140(1)}{\pico\coulomb} electron beams with mean and standard deviation energy $\approx\qty[parse-numbers = false]{609  \pm 2(stat) (12)(syst)}{\mega\electronvolt}$ and shot-to-shot standard deviation \SI[separate-uncertainty]{40(1)}{\mega\electronvolt}, respectively (standard error given). Electrons with energies $<\SI{300}{\mega\electronvolt}$ could not be measured.
A radial source size of $<\SI[separate-uncertainty]{0.7(1)}{\micro\metre}$ was assumed, in line with previous measurements~\citemethod{Kneip_2011}\citemethod{Schnell_2012}. The FWHM energy-dependent electron beam divergence, $\theta_D$, measured along the axis transverse to the dispersion plane by the LANEX screens which measured the electron spectrum, was $(b_1-b_2\sqrt{\gamma m_e[\SI[]{}{\giga\electronvolt}]})$, where $b_1=1.30^{+0.26}_{-0.19}\SI[]{}{\milli\radian}$, $b_2={0.26^{+0.24}_{-0.28}}\SI{}{\milli\radian\per\giga\electronvolt\tothe{1/2}}$. The axial symmetry of the electron beam divergence was confirmed using linear Thomson scattering~\citemethod{Gerstmayr_2020}.

The colliding laser pulse was focused at the rear of the gas jet by an \textit{f}/2 parabola with a \SI[]{25.4}{\milli\metre} on-axis hole. The laser intensity profile had transverse FWHM~\SI[separate-uncertainty]{2.5(2)}{\micro\metre}$\times$\SI[separate-uncertainty]{2.1(1)}{\micro\metre} and FWHM duration~\SI[separate-uncertainty]{45.0(25)}{\femto\second}. Due to energy losses in the laser system, including the on-axis hole in the \textit{f}/2 parabola, the energy on-target was \SI[separate-uncertainty]{6.13(2)}{\joule}, yielding a peak $a_0=21.4\pm1.8$.


\subsection{Spatial and Temporal Overlap of Laser pulses}\label{subsec:laser_timing_and_alignment}

Accessing collision $a_0\gtrsim10$ required $\lesssim\SI{1}{\micro\metre}$ and $\lesssim\SI{10}{\femto\second}$ precision spatio-temporal overlap of the electron beam and laser focus. This required a careful alignment procedure, detailed below.

A micron knife-edge \SI[]{90}{\degree} prism, imaged using a $\times 10$ microscope objective, was used to overlap the two laser pulses spatially and temporally, where the latter was achieved using spatial interferometry, as demonstrated previously~\citemethod{Cole_2018}\citemethod{Corvan_2016}. By optimising the contrast of the interference pattern, the two laser pulses were synchronised to within $\pm$\SI[]{10}{\femto\second}. To time the colliding pulse, two additional effects needed to be corrected for, namely the reduced non-linear group velocity of the wakefield-driver laser pulse in the plasma and the longitudinal displacement of the electron beam from the wakefield driver by $N$ plasma wavelengths, where $N=\frac{1}{2}$ for an electron beam travelling at the dephasing limit with velocity close to $c$. Thus, the longitudinal collision position was shifted closer to the gas jet by $\delta_z$,
\begin{equation}
\delta_z=\frac{3d}{4}\frac{n_e}{n_{c}}+N\frac{\lambda_L}{2}\sqrt{\frac{n_{c}}{n_e}}
\end{equation}
where $n_e$ and $n_{c}=\frac{\epsilon_0 m_e\omega_L^2}{e^2}$ are the plasma and critical densities and $d$ is the distance from the upstream edge of the plasma to the injection point. 

Fluctuations in ambient temperature altered the temporal and spatial alignment between the electron beam and colliding laser over the course of shooting. Thus, implementing timing and pointing stabilisation was key to obtaining a high number of successful collisions.

Once timed and aligned, references for optimal timing and alignment were taken using a spectral interferometer in the laser area and a diagnostic of the \textit{f}/40 beam pointing, respectively. Long-term drifts in spatial and temporal alignment could be corrected for by adjusting the tip and tilt of a mirror in the \textit{f}/40 beamline, and by altering the path length of one of the laser arms, respectively. 

This was implemented using an automated feedback loop. Thus, the remaining misalignment between the electron beam and the colliding laser resulted from shot-to-shot variation in the beam paths and the laser pointing due to vibrations. The temporal jitter between the two laser arms was assumed to be normally distributed with standard deviation $\pm$\SI[]{30}{\femto\second}. An additional source of uncertainty in the timing between the electron beam and colliding laser stems from the unknown value of $d$, which we assume to be uniformly distributed with lower and upper bounds \SI[]{0}{\milli\metre} and \SI[]{10}{\milli\metre}, respectively. To correct for $\delta_z$, the path between the two beams was reduced by \SI[]{20}{\femto\second}. Thus, the offset in timing between the two beams, $\Delta t\approx\frac{\delta_z}{c}$, $\SI[]{2.7}{\femto\second}\leq \Delta t\leq\SI[]{45.8}{\femto\second}$.
The standard deviations of the radial positions of the colliding laser and the electron beam due to pointing variations were measured to be \SI[separate-uncertainty]{0.53(26)}{\micro\metre} and \SI[separate-uncertainty]{17.5(5)}{\micro\metre} respectively, which correspond to a standard deviation in transverse alignment of \SI[separate-uncertainty]{17.5(6)}{\micro\metre}.



\subsection{Gamma Radiation Diagnostics}\label{subsec:gamma_stack_calib_retrieval}

The angular distribution of gamma radiation was measured using a \SI[parse-numbers=false]{50\times50\times10}{\milli\metre} profile screen consisting of \SI[parse-numbers=false]{1 \times 1 \times 10}{\milli\metre} CsI(Tl) crystals separated by \SI[]{0.2}{\milli\metre} titanium oxide spacers. Dimensions have the format (horizontal $\times$ vertical $\times$ depth). The front of the profile screen was coated with \SI[]{0.5}{\milli\metre} titanium oxide. The profile screen was placed outside the vacuum chamber, \SI[separate-uncertainty]{2244(4)}{\milli\metre} from the interaction, and was imaged using a cooled 16-bit CCD camera.

The energy deposition of gamma photons was measured using a \SI[parse-numbers=false]{50 \times 50 \times 150}{\milli\metre} dual-axis CsI(Tl) scintillator, comprised of alternating layers of horizontally and vertically oriented \SI[parse-numbers=false]{5 \times 5 \times 50}{\milli\metre} CsI(Tl) crystals which were held in place by a 3D printed \SI[]{1}{\milli\metre} nylon frame and separated by \SI[]{1}{\milli\metre} rubber spacers to prevent light leakage between crystals. Two cooled 16-bit CCD cameras imaged the scintillation light from above and laterally. The calorimeter was placed outside the vacuum chamber, \SI[separate-uncertainty]{3570(3)}{\milli\metre} from the interaction. 

Geant4~\citemethod{Allison_2016}\citemethod{Agostinelli_2003} simulations were used to obtain the energy deposition in the CsI photon diagnostics~\citemethod{Gerstmayr_2020} as a function of incident photon energy as demonstrated in Behm \textit{et al.}~\citemethod{Behm_2018}. The maximum and minimum photon energies used to calculate energy deposition were \SI{0.01}{\mega\electronvolt} and \SI{1}{\giga\electronvolt}, respectively. These simulations included the chamber geometry, large objects inside the chamber such as the dipole magnets and all materials placed in the beam path, including a \SI[]{1}{\milli\metre} alumina laser block, a \SI[]{25}{\micro\metre} Kapton window with a \SI[]{375}{\micro\metre} Kevlar backing sheet and a \SI[]{25}{\micro\metre} aluminium foil. Variations in the scintillation efficiency of the crystals and in the efficiency of the imaging system were characterised and subsequently compensated for by comparing the measured and simulated response of the calorimeter to bremsstrahlung generated by an electron beam propagating through a \SI[separate-uncertainty]{1.5(1)}{\milli\metre} PTFE target with radiation length much less than the radiation length of a $\SI[]{1}{\giga\electronvolt}$ electron beam.

The ICS spectrum, $S_{ICS}$, has the characteristic shape 
\begin{equation}
\label{eqn:gamma_spec}
S_{ICS}=A\left(\frac{E_\gamma}{E_{\gamma c}}\right)^{-\frac{2}{3}}e^{-\left(\frac{E_\gamma}{E_{\gamma c}}\right)}
\end{equation}
where $A$ and $E_{\gamma c}$ represent photon number and the critical energy of the spectrum, respectively. Bayesian inference was used to obtain the values of $A$ and $E_{\gamma c}$ for which the energy deposition calculated using equation~\ref{eqn:gamma_spec} fitted the measured energy deposition.

\subsection{Methods for Frequentist Analysis}\label{subsec:shot_selection}


Following the approach employed by Cole \textit{et. al.}~\citemethod{Cole_2018}, the photon yield measured by the gamma profile diagnostic was used to identify hits and nulls. The total yield measured by the profile screen, $Y_{\gamma}$, is expected to scale with the total charge, $Q$ and expected value of $\gamma^2$, $\langle\gamma^2\rangle$ of the electron beam:
\begin{equation}
Y_{\gamma}=(C_{\rm{ICS}}a_0^2 +C_{\rm{BKG}}) Q\langle\gamma^2\rangle
\end{equation}
where the first and second terms describe the contributions of inverse Compton scattering (ICS) and background (e.g. due to bremsstrahlung) to the total yield, respectively, and $C_{\rm{ICS}}$, $C_{\rm{BKG}}$, are scaling constants. The scaling for inverse Compton scattering with photon yield holds for $\gamma a_0^2<5.5\times10^{5}$~\citemethod{Thomas_2012}. Multiple sets of shots in which the counter-propagating laser was not fired were taken to obtain the characteristic background scaling with $Q\langle\gamma^2\rangle$. Misses and hits, classified as shots which produced yields within $1\sigma$ and above $3\sigma$ of the background scaling respectively, are shown in figure~\ref{fig:gp_shot_sel}.

In this work, we present a controlled experiment in which the only difference between hits and nulls is the presence or absence of a collision between a laser and an electron beam, respectively. Systematic changes in electron beam properties and the effect of background radiation mechanisms were mitigated and accounted for if necessary.

Automated correction of the spatial and temporal overlap of the wakefield drive laser and the colliding laser was implemented to minimise long-term drifts. Hits and nulls were interleaved to ensure null electron beams were representative of hits.

Background radiation mechanisms, namely betatron and bremsstrahlung, were characterised using laser-off shots, and yielded a positive scaling of photon yield with $Q\langle\gamma^2\rangle$. Distributions of the mean electron beam energy, $\langle E\rangle$ and height of its spectral peak above the $70^{th}$ percentile energy, $P_{70}$, were obtained for hits and nulls.

All radiation reaction models predict a reduction in $\langle E\rangle$ and $P_{70}$ (the latter is predicted for all models if the electron beam interacts with a range of laser intensities), and a photon yield which increases with decreasing $P_{70}$ and decreasing $\langle E\rangle$. By contrast, background radiation mechanisms produce the opposite; a positive scaling between photon yield and $Q\langle\gamma^2\rangle$. Given that the only difference between hits and nulls is the presence or absence of a collision between the electron beam and laser, the simultaneous observation of all of the above signatures of radiation reaction signatures for hits and their absence for nulls, constitutes definitive evidence of radiation reaction.



\subsection{Analytical Methods}\label{sec:ana_methods}


\subsubsection{Free and Fixed Parameter Selection, Bayesian Inference Test Cases}\label{sec:BI_excl_param_summary}

 A full, detailed account of the implementation and testing of the Bayesian inference framework and the forward models used therein is provided in~\citemethod{Los_2024}.

Several parameters (including electron beam source size and chirp, laser duration, transverse offset, etc) have been assigned fixed values in the forward model (i.e. are not inferred). Free and fixed parameters were chosen in accordance with the following criteria:
\begin{itemize}
\item The expected effect of variation. This incorporates both the probability of parameter variation by a given amount and the impact of this variation on the post-collision electron and photon spectra. Parameters fixed due to their small expected effect include laser duration and focal spot size and electron beam chirp.

\item Shot selection. The ten shots which produced the highest CsI profile screen yields, normalised to $Q\langle\gamma^2\rangle$, were analysed using the Bayesian framework. As the laser intensity decreases most steeply with transverse (rather than longitudinal) misalignment, by analysing the highest yield shots the probability of a large transverse offset is reduced. 

\item Degeneracy. If changes in two (or more) collision parameters engender similar alterations in the post-collision observables, it is possible to fix one of these parameters and vary the second to reproduce the effect of the fixed parameter. For example, if the electron beam has finite divergence, varying the longitudinal position of the collision alters the size of the electron beam at the collision. This produces post-collision observables similar to those obtained by varying the electron beam source size. Degeneracy allows changes in the electron beam source size, divergence and transverse offset from the laser focus to be compensated by free parameters (laser energy, longitudinal offset of the collision from focus, electron beam duration).
The laser energy was chosen as a free parameter to enable the Bayesian inference procedure to tackle shot-to-shot variations therein. 
\end{itemize}
The laser, electron and collision parameters which were measured, estimated or inferred based on previous measurements, are summarised alongside their assigned values in the forward models in tables~\ref{tab:exp_laser_params},~\ref{tab:exp_ebeam_params} and~\ref{tab:coll_params}, respectively in supplementary material~\ref{subsec:coll_params}.


\subsubsection{Bayesian inference implementation and testing}

The Bayesian inference procedures used the Markov chain Monte-Carlo (MCMC)~\citemethod{Brooks_2011} from the python package emcee. 
Lack of on-shot parameter measurements necessitated broad priors~\citemethod{Los_2024}.

Extensive testing of the Bayesian analysis~\citemethod{Los_2024}, which included inferences on synthetic data where the fixed collision parameters (e.g. $Z_d$) were assigned different values from the forward models, revealed that for all test cases, the inference procedure yielded weak evidence favouring the correct model; none of the test cases yielded false positives. This is a clear demonstration of the accuracy of the model selection capabilities of the Bayesian framework we have developed. In all test cases, the highest performing model(s) inferred $\langle \tilde{a}_0\rangle$ and $\langle\eta\rangle$ within $1\sigma$ of the input (correct) value. This shows the analysis infers the physical parameters governing the collision accurately to first order, in spite of the simplifying assumptions made.

\subsubsection{Bayes factors}

The Bayes factor, $P_X/P_Y$, for models $X$ and $Y$, used to perform model comparison, is defined as the integral over the marginalised posterior/likelihood
\begin{equation}
\label{eqn:bic}
\frac{P_X}{P_Y}=\frac{\int p(\phi_X|M_X)p(D|\phi_X, M_X)d\phi_X}{\int p(\phi_Y|M_Y)p(D|\phi_Y, M_Y)d\phi_Y}
\end{equation}
where $\phi_X$, $\phi_Y$ are the parameter vectors which characterise models $M_X$ and $M_Y$, respectively. The integrals in equation~\ref{eqn:bic} do not have analytic solutions and are challenging to compute numerically due to the complex shape of the posterior distribution. Therefore, the Bayes factor was approximated using leave-one-out cross-validation with Pareto-smoothed importance sampling (LOO-PSIS)~\citemethod{Vehtari_2017} available from the python package arviz~\citemethod{arviz_2019}.
As the inferred parameters and hence the posterior probabilities for each shot are independent, their product yields the total Bayes factor.

\subsubsection{Bootstrapping}\label{Sec:bootstrapping}

Bootstrapping, chosen for its applicability to non-normal distributions, was used to compute the errors on population means of the hit and null distributions for $\langle E\rangle$ and $P_{70}$, shown in figure~\ref{fig:eloss_hists}. Bootstrapping was also used to compute the mean and standard error for the null scaling of photon counts measured by the profile screen with $Q\langle \gamma^2 \rangle$, illustrated in figure~\ref{fig:gp_shot_sel}. In each instance where bootstrapping was employed in the analysis, a sample size equal to that of the dataset was used, and the data was re-sampled 10000 times. The robustness of the bootstrapping analysis was verified by increasing the number of re-samples by factor of 100, which did not significantly affect the results. The random.randint package from the numpy library in python was used to perform sampling.


\clearpage

\section{Data Availability}

The authors declare that all data supporting the findings of this study are available within the article and its Supplementary Information files or from the corresponding author upon reasonable request.

\section{Code Availability}

The authors declare that all code supporting the findings of this study are available on Github at https://github.com/ELos385/RadiationReaction/tree/main and on Code Ocean at https://codeocean.com/capsule/7757204/tree.

\clearpage

\bibliographystylemethod{unsrt}
\nocite{*}
\bibliographymethod{method}

\clearpage

\section{Acknowledgements}

The authors would like to kindly thank Dr. Roberto Fumagalli for his insightful comments regarding the manuscript revisions.

\section{Ethics declarations}
\subsection{Competing interests}
The authors declare no competing financial interests.


\backmatter



\appendix
\renewcommand{\thesection}{\Alph{section}}
\renewcommand{\thesubsection}{\Alph{section}.\arabic{subsection}}


\section{Supplementary Note}\label{sup:BI_excl_param_summary}
\subsection*{Collision parameters}\label{subsec:coll_params}

\begin{table}[!ht]
\centering
\begin{tabular}{ccccp{9cm}}
\toprule
Laser parameters &Experiment & Value in forward model\\
\toprule
Energy on target (\SI[]{}{\joule})  &  $6.13\pm 0.02$ & Free parameter\\
FWHM transverse waist (\SI[]{}{\micro\meter}) & $\SI[separate-uncertainty]{2.5(2)}{}\times\SI[separate-uncertainty]{2.1(1)}{}$ & 2.47\\
FWHM duration (\SI[]{}{\femto\second}) & $45\pm3$ & 45\\
\bottomrule
\end{tabular}
\caption{Measured laser parameters.\label{tab:exp_laser_params}}
\end{table}

\begin{table}[!ht]
\centering
\begin{tabular}{cccccp{9cm}}
\toprule
Electron beam property & Experiment&Value in forward model\\ 
\toprule
Duration (standard deviation) (\SI[]{}{\femto\second}) & \SI[separate-uncertainty]{14(14)}{} & Free parameter\\
Transverse source size& \SI[separate-uncertainty]{0.68(13)}{} & 0.68\\
(standard deviation) (\SI[]{}{\micro\meter})&&\\
Distance from end of gas jet to& 0.0&0.0\\
electron beam initial position (\SI[]{}{\milli\meter}) &&\\
Total electron charge (\SI[]{}{\pico\coulomb})& \SI[separate-uncertainty]{140(1)}{}& Normalised\\
FWHM divergence (\SI[]{}{\milli\radian}) & $(b_1-b_2\sqrt{\gamma m_e[\SI[]{}{\giga\electronvolt}]})$& $(b_1-b_2\sqrt{\gamma m_e[\SI[]{}{\giga\electronvolt}]})$\\
\bottomrule
\end{tabular}
\caption{Measured or estimated electron beam parameters. The electron beam source size has been estimated from previous measurements~\protect\citesupp{Schnell_2012}, while the electron beam duration was obtained from particle-in-cell simulations using the code FBPIC~\protect\citesupp{Lehe_2016}. The constants $b_1=1.30^{+0.26}_{-0.19}~\SI[]{}{\milli\radian}$, $b_2={0.26^{+0.24}_{-0.28}}~\SI{}{\milli\radian\per\giga\electronvolt\tothe{1/2}}$. \label{tab:exp_ebeam_params}}
\end{table}

\begin{table}[!ht]
\centering
\begin{tabular}{ccccp{9cm}}
\toprule
Collision parameters &Experiment & Value in forward model\\
\toprule
Transverse displacement of&  $\pm17.5$ & 0.0\\
collision from focus (\SI[]{}{\micro\meter}) & &\\
Temporal displacement of&  $\pm N(0, 30)$ & Free parameter\\
collision from focus (\SI[]{}{\femto\second}) & $+ U(3, 46)$&\\
\bottomrule
\end{tabular}
\caption{The expected transverse and temporal alignment of the electron beam and the colliding laser and the expected shot-to-shot jitter in the above parameters. $U$ and $N$ denote uniform and normal distributions, respectively.\label{tab:coll_params}}
\end{table}

\clearpage

\section{Supplementary Note}\label{sup:extra_BI_results}
\subsection*{Bayesian Inference Additional Results}\label{subsec:extra_BI_result}

\begin{figure}[ht!]
\centering
\begin{subfigure}[t]{0.99\textwidth}
{\begin{overpic}[width=0.99\linewidth, trim={0.1cm, 0.28cm, 0.25cm, 0.26cm},clip]{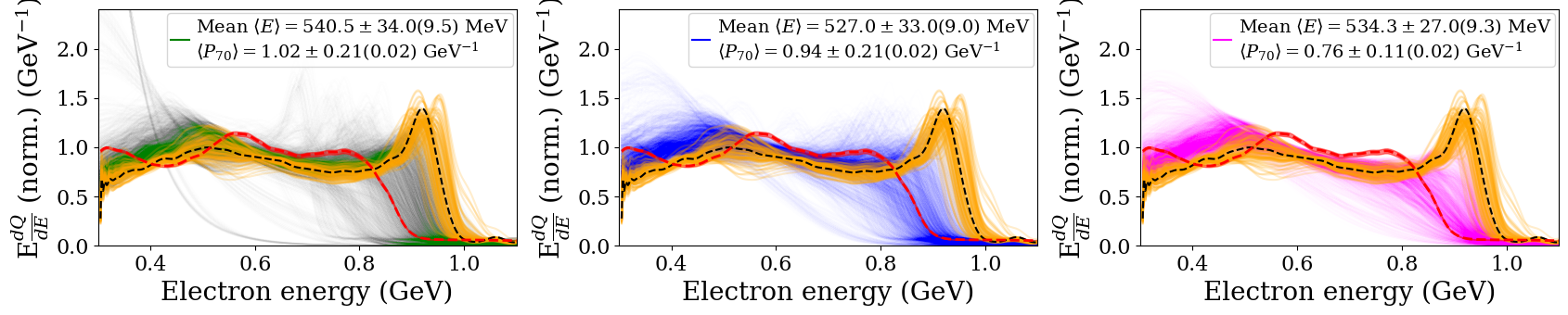}
\put(-2,18){a)}
\put(14,20.5){Classical}
    \put(40.5,20.5){Quantum-continuous}
	\put(74.5, 20.5){Quantum-stochastic}
\end{overpic}}
\end{subfigure}
\vspace{0.15em}
\\
 \begin{subfigure}[t]{0.99\textwidth}
{\begin{overpic}[width=0.99\linewidth, trim={0.1cm, 0.25cm, 0.21cm, 0.2cm},clip]{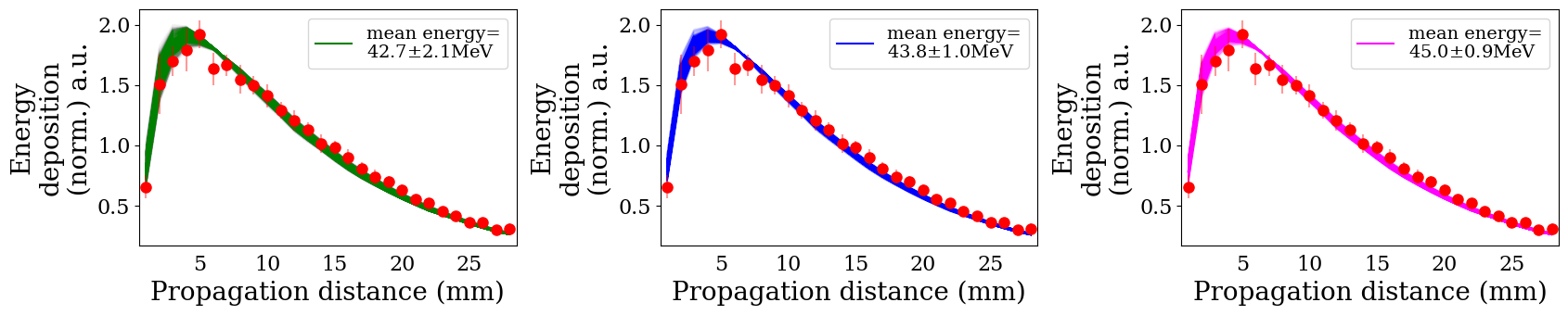}
\put(-2,18){b)}
\end{overpic}}
\end{subfigure}
\caption{\textbf{Bayesian inference results for shot 5.}  Similar to figure~\ref{fig:post-coll_espec_no_chirp_1}. The distribution of $\tilde{a}_{0}$ inferred by the classical, quantum-continuous and quantum-stochastic models had mean and standard deviation, $\langle \tilde{a}_{0}\rangle=7.5\pm1.2$ and $\sigma_{a_0}=1.9\pm0.5$, $\langle \tilde{a}_{0}\rangle=8.9\pm1.2$ and $\sigma_{a_0}=2.8\pm0.6$ and $\langle \tilde{a}_{0}\rangle=9.4\pm1.4$ and $\sigma_{a_0}=0.9\pm0.2$, respectively. a) The measured post-collision electron spectra, with mean $\langle E\rangle=\qty[parse-numbers = false]{(547.8 \pm 0.0 (9.7))}{\mega\electronvolt}$ and $\langle P_{70} \rangle=\qty[parse-numbers = false]{0.88\pm 0.00 (0.01)}{\per\giga\electronvolt}$ and pre-collision spectra predicted by the neural network, for which mean $\langle E\rangle=\qty[parse-numbers = false]{590.5 \pm 9.3 (11.3)}{\mega\electronvolt}$ and $\langle P_{70} \rangle=\qty[parse-numbers = false]{1.33\pm 0.09(0.02)}{\per\giga\electronvolt}$, are shown alongside the predicted post-collision electron spectra. b) The measured post-collision photon spectrum, with mean energy $\SI[separate-uncertainty]{62.4(66)}{\mega\electronvolt}$ and predicted photon spectra.
\label{fig:post-coll_spec_no_chirp_0}}
\end{figure}

Results of the Bayesian inference procedure applied to nine shots, ordered by the corresponding normalised gamma profile signal with the brightest shots first.
In figure~\ref{fig:post-coll_spec_no_chirp_0}, the low energy peak in the measured post-collision (red) electron spectrum is not present in the pre-collision spectrum (orange, median shown in black) predicted by the neural network, likely due to a deficiency of spectra with this feature in the training data.


\begin{figure}[ht!]
\centering
\begin{subfigure}[t]{0.99\textwidth}
{\begin{overpic}[width=0.99\linewidth, trim={0.1cm, 0.20cm, 0.20cm, 0.22cm},clip]{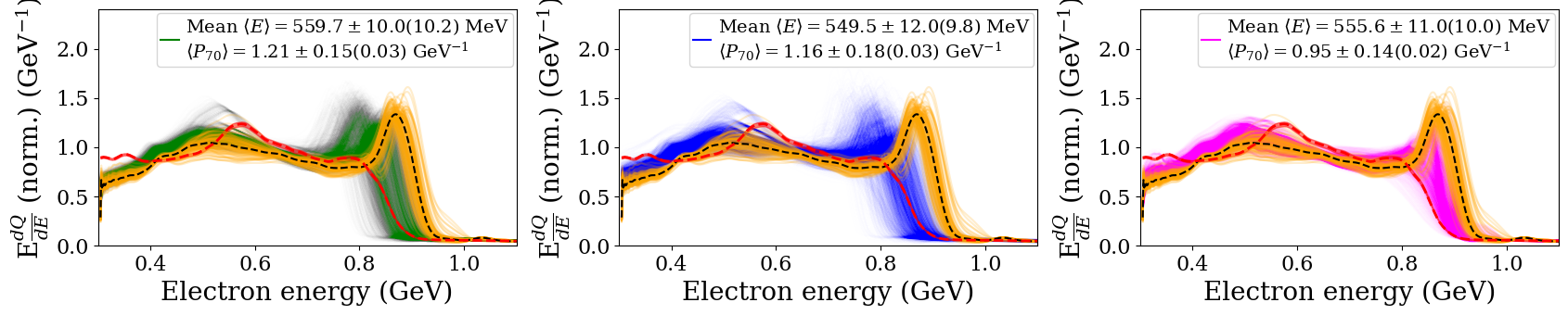}
\put(-2,18){a)}
\put(14,20.5){Classical}
    \put(40.5,20.5){Quantum-continuous}
	\put(74.5, 20.5){Quantum-stochastic}
\end{overpic}}
\end{subfigure}
\vspace{0.15em}
\\
 \begin{subfigure}[t]{0.99\textwidth}
{\begin{overpic}[width=0.99\linewidth, trim={0.1cm, 0.25cm, 0.20cm, 0.2cm},clip]{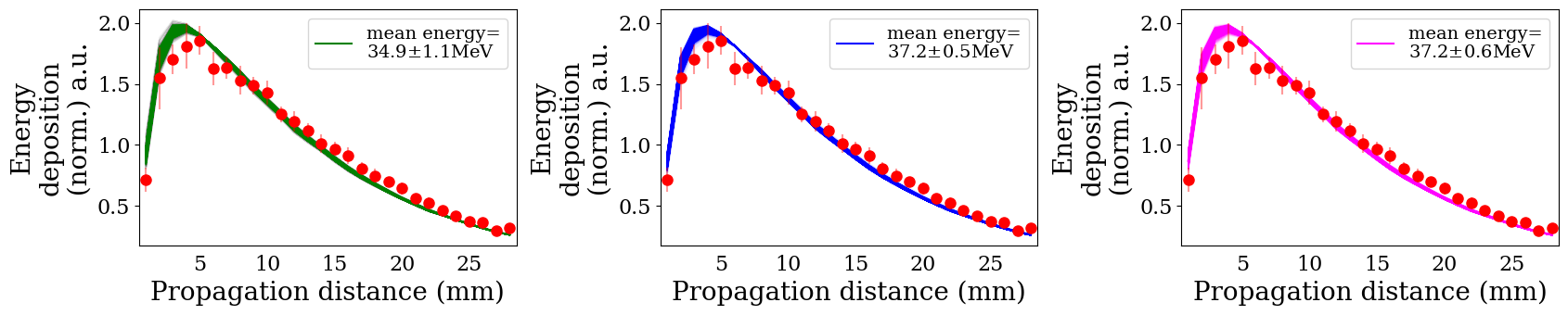}
\put(-2,18){b)}
\end{overpic}}
\end{subfigure}
\caption{\textbf{Bayesian inference results for shot 4.} Similar to figure~\ref{fig:post-coll_espec_no_chirp_1}. The distribution of $\tilde{a}_{0}$ inferred by the classical, quantum-continuous and quantum-stochastic models had mean and standard deviation, $\langle \tilde{a}_{0}\rangle=6.2\pm1.1$ and $\sigma_{a_0}=1.2\pm0.3$, $\langle \tilde{a}_{0}\rangle=6.7\pm0.8$ and $\sigma_{a_0}=0.4\pm0.1$ and $\langle \tilde{a}_{0}\rangle=6.7\pm0.9$ and $\sigma_{a_0}=0.5\pm0.1$, respectively. a) The measured post-collision electron spectrum, with $\langle E \rangle=\qty[parse-numbers = false]{545.5 \pm0.0 (9.7)}{\mega\electronvolt}$ and $P_{70}=\qty[parse-numbers = false]{0.88 \pm 0.00 (0.01)}{\per\giga\electronvolt}$. The neural network predictions for the pre-collision electron spectrum have $\langle E \rangle=\qty[parse-numbers = false]{580.2 \pm 4.4 (10.9)}{\mega\electronvolt}$ and $P_{70}=\qty[parse-numbers = false]{1.38\pm 0.07(0.02)}{\per\giga\electronvolt}$. b) The measured photon spectrum, with mean energy \SI[separate-uncertainty]{72(23)}{\mega\electronvolt}.
\label{fig:post-coll_spec_no_chirp_7}}
\end{figure}

\begin{figure}[ht!]
\centering
\begin{subfigure}[t]{0.99\textwidth}
{\begin{overpic}[width=0.99\linewidth, trim={0.1cm, 0.20cm, 0.20cm, 0.22cm},clip]{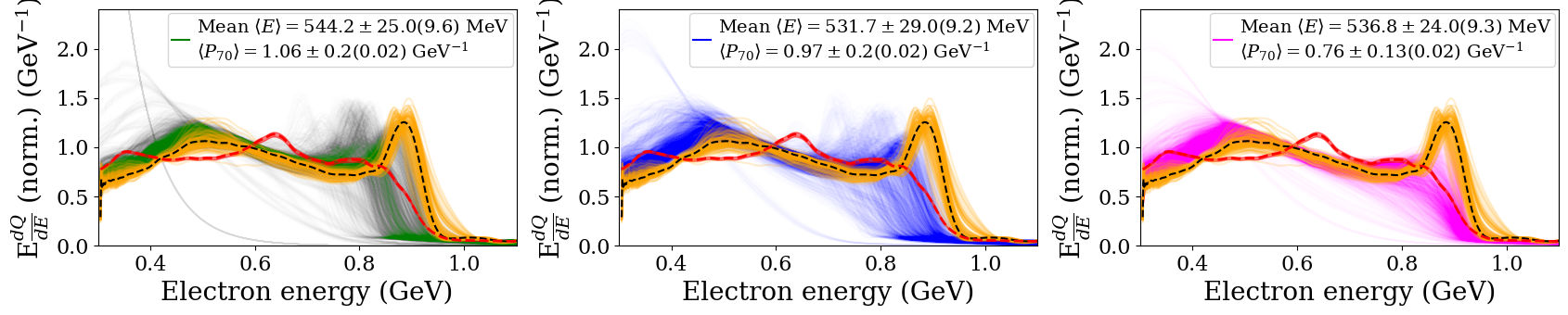}
\put(-2,18){a)}
\put(14,20.5){Classical}
    \put(40.5,20.5){Quantum-continuous}
	\put(74.5, 20.5){Quantum-stochastic}
\end{overpic}}
\end{subfigure}
\vspace{0.15em}
\\
 \begin{subfigure}[t]{0.99\textwidth}
{\begin{overpic}[width=0.99\linewidth, trim={0.1cm, 0.25cm, 0.20cm, 0.2cm},clip]{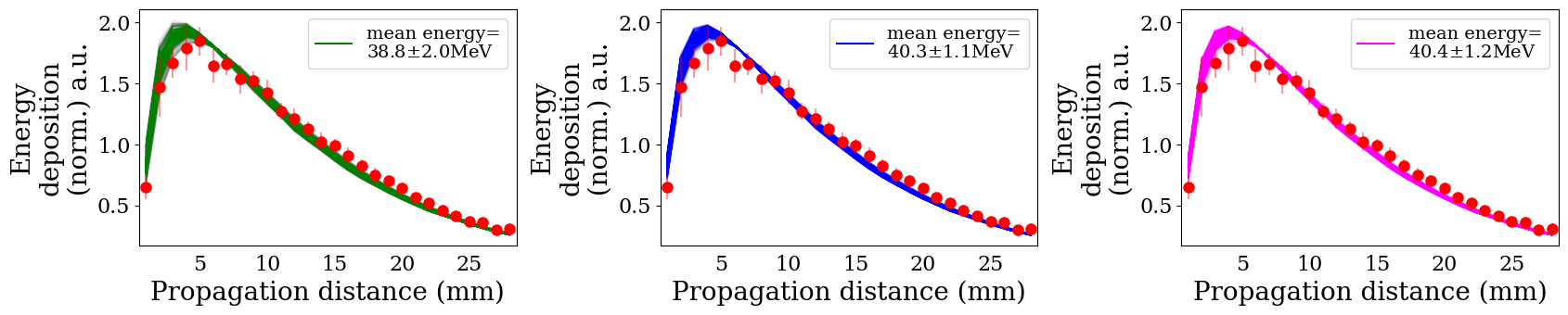}
\put(-2,18){b)}
\end{overpic}}
\end{subfigure}
\caption{\textbf{Bayesian inference results for shot 8.} Similar to figure~\ref{fig:post-coll_espec_no_chirp_1}. The distribution of $\tilde{a}_{0}$ inferred by the classical, quantum-continuous and quantum-stochastic models had mean and standard deviation, $\langle \tilde{a}_{0}\rangle=6.3\pm1.0$ and $\sigma_{a_0}=0.3\pm0.1$, $\langle \tilde{a}_{0}\rangle=7.7\pm1.1$ and $\sigma_{a_0}=0.6\pm0.2$ and $\langle \tilde{a}_{0}\rangle=7.7\pm1.0$ and $\sigma_{a_0}=0.5\pm0.1$, respectively. 
a) The measured post-collision electron spectrum, with $\langle E \rangle=\qty[parse-numbers = false]{557.1 \pm 0.0 (10.1)}{\mega\electronvolt}$ and $P_{70}=\qty[parse-numbers = false]{0.81\pm 0.00 (0.01)}{\per\giga\electronvolt}$. The neural network predictions for the pre-collision electron spectrum have $\langle E \rangle=\qty[parse-numbers = false]{579.4 \pm 5.9 (10.9)}{\mega\electronvolt}$ and $P_{70}=\qty[parse-numbers = false]{1.3 \pm 0.06 (0.02)}{\per\giga\electronvolt}$. b) The measured photon spectrum, with mean energy \SI[separate-uncertainty]{67(6)}{\mega\electronvolt}.
\label{fig:post-coll_spec_no_chirp_6}}
\end{figure}

\begin{figure}[ht!]
\centering
\begin{subfigure}[t]{0.99\textwidth}
{\begin{overpic}[width=0.99\linewidth, trim={0.1cm, 0.20cm, 0.20cm, 0.22cm},clip]{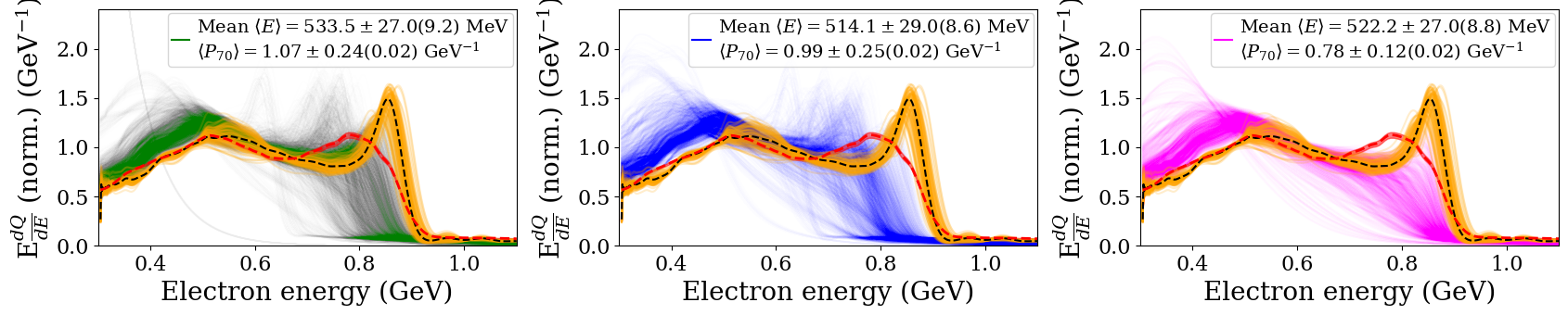}
\put(-2,18){a)}
\put(14,20.5){Classical}
    \put(40.5,20.5){Quantum-continuous}
	\put(74.5, 20.5){Quantum-stochastic}
\end{overpic}}
\end{subfigure}
\vspace{0.15em}
\\
 \begin{subfigure}[t]{0.99\textwidth}
{\begin{overpic}[width=0.99\linewidth, trim={0.1cm, 0.25cm, 0.20cm, 0.2cm},clip]{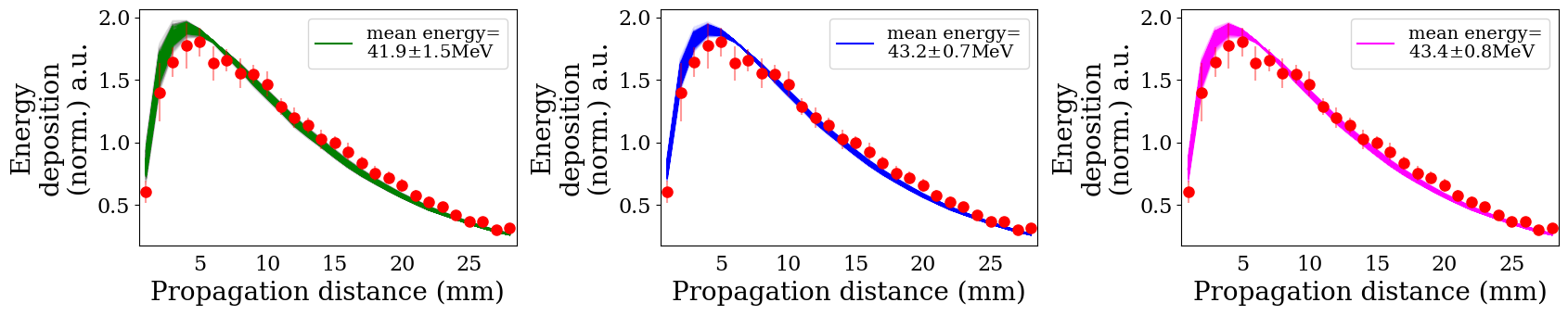}
\put(-2,18){b)}
\end{overpic}}
\end{subfigure}
\caption{\textbf{Bayesian inference results for shot 9.} Similar to figure~\ref{fig:post-coll_espec_no_chirp_1}. The distribution of $\tilde{a}_{0}$ inferred by the classical, quantum-continuous and quantum-stochastic models had mean and standard deviation, $\langle \tilde{a}_{0}\rangle=7.1\pm1.1$ and $\sigma_{a_0}=0.3\pm1.0$, $\langle \tilde{a}_{0}\rangle=9.3\pm1.0$ and $\sigma_{a_0}=2.1\pm0.5$ and $\langle \tilde{a}_{0}\rangle=9.1\pm1.2$ and $\sigma_{a_0}=0.7\pm0.1$, respectively. a) The measured post-collision electron spectrum, with $\langle E \rangle=\qty[parse-numbers = false]{573.8 \pm 0.0 (10.7)}{\mega\electronvolt}$ and $P_{70}=\qty[parse-numbers = false]{0.93 \pm 0.00 (0.01)}{\per\giga\electronvolt}$. The neural network predictions for the pre-collision electron spectrum have $\langle E \rangle=\qty[parse-numbers = false]{578.6 \pm 4.0 (10.9)}{\mega\electronvolt}$ and $P_{70}=\qty[parse-numbers = false]{1.48 \pm 0.1 (0.02)}{\per\giga\electronvolt}$. b) The measured photon spectrum, with mean energy \SI[separate-uncertainty]{73(7)}{\mega\electronvolt}.
\label{fig:post-coll_spec_no_chirp_5}}
\end{figure}

\begin{figure}[ht!]
\centering
\begin{subfigure}[t]{0.99\textwidth}
{\begin{overpic}[width=0.99\linewidth, trim={0.1cm, 0.20cm, 0.20cm, 0.22cm},clip]{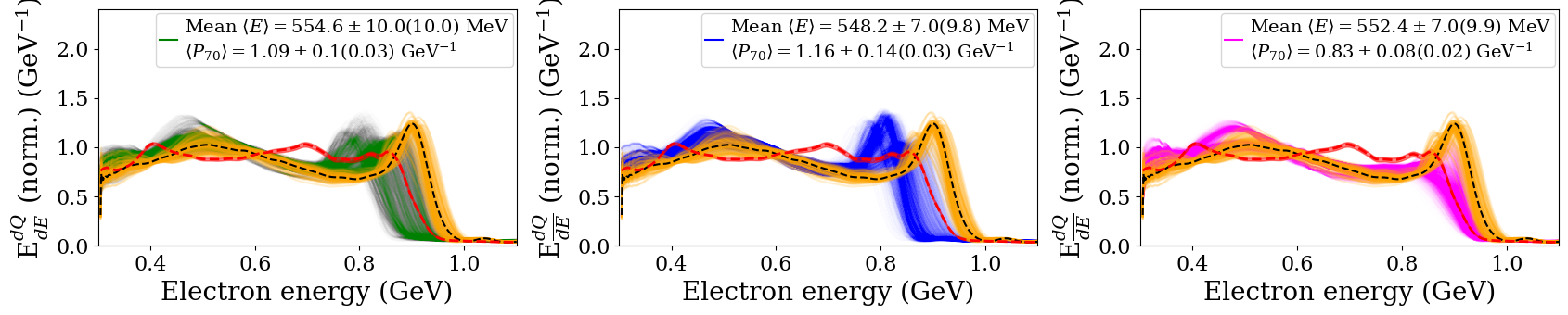}
\put(-2,18){a)}
\put(14,20.5){Classical}
    \put(40.5,20.5){Quantum-continuous}
	\put(74.5, 20.5){Quantum-stochastic}
\end{overpic}}
\end{subfigure}
\vspace{0.15em}
\\
 \begin{subfigure}[t]{0.99\textwidth}
{\begin{overpic}[width=0.99\linewidth, trim={0.1cm, 0.25cm, 0.20cm, 0.2cm},clip]{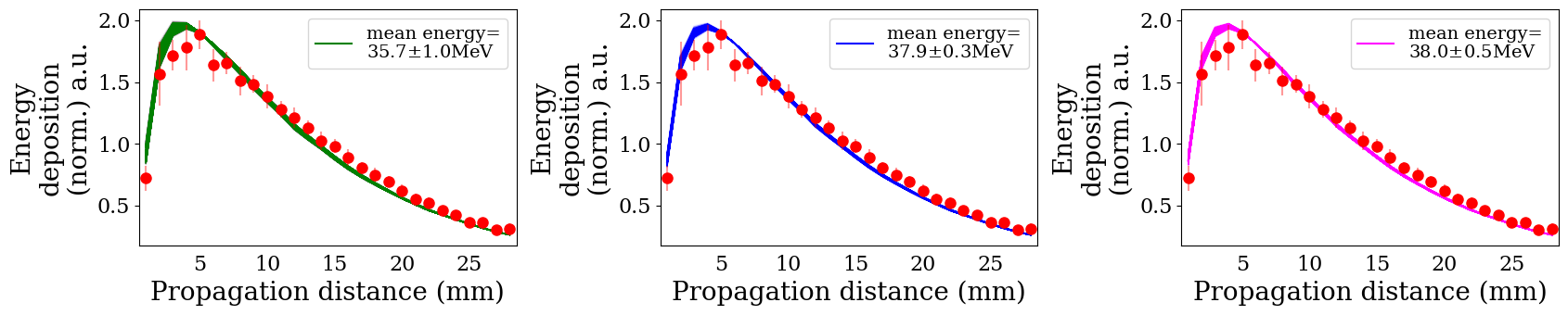}
\put(-2,18){b)}
\end{overpic}}
\end{subfigure}
\caption{\textbf{Bayesian inference results for shot 6.} Similar to figure~\ref{fig:post-coll_espec_no_chirp_1}. The distribution of $\tilde{a}_{0}$ inferred by the classical, quantum-continuous and quantum-stochastic models had mean and standard deviation, $\langle \tilde{a}_{0}\rangle=6.5\pm1.1$ and $\sigma_{a_0}=1.3\pm0.2$, $\langle \tilde{a}_{0}\rangle=8.1\pm1.1$ and $\sigma_{a_0}=0.9\pm0.2$ and $\langle \tilde{a}_{0}\rangle=6.9\pm0.8$ and $\sigma_{a_0}=0.5\pm0.1$, respectively. a) The measured post-collision electron spectrum, with $\langle E \rangle=\qty[parse-numbers = false]{562.3 \pm 0.0 (10.3)}{\mega\electronvolt}$ and $P_{70}=\qty[parse-numbers = false]{0.96 \pm 0.00 (0.01)}{\per\giga\electronvolt}$. The neural network predictions for the pre-collision electron spectrum have $\langle E \rangle=\qty[parse-numbers = false]{575.3 \pm 5.9 (10.7)}{\mega\electronvolt}$ and $P_{70}=\qty[parse-numbers = false]{1.22 \pm 0.06 (0.02)}{\per\giga\electronvolt}$. b) The measured photon spectrum, with mean energy \SI[separate-uncertainty]{62(7)}{\mega\electronvolt}. 
\label{fig:post-coll_spec_no_chirp_4}}
\end{figure}

\begin{figure}[ht!]
\centering
\begin{subfigure}[t]{0.99\textwidth}
{\begin{overpic}[width=0.99\linewidth, trim={0.1cm, 0.20cm, 0.20cm, 0.22cm},clip]{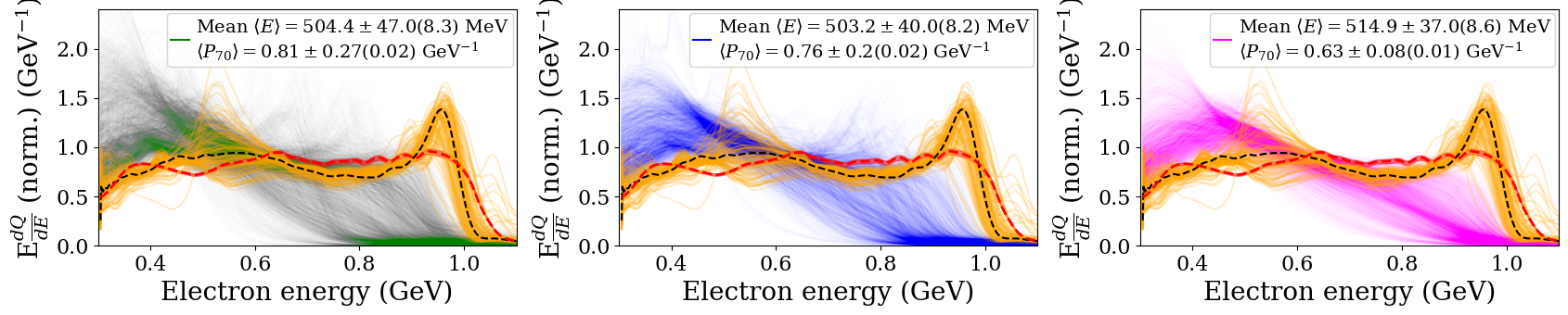}
\put(-2,18){a)}
\put(14,20.5){Classical}
    \put(40.5,20.5){Quantum-continuous}
	\put(74.5, 20.5){Quantum-stochastic}
\end{overpic}}
\end{subfigure}
\vspace{0.15em}
\\
 \begin{subfigure}[t]{0.99\textwidth}
{\begin{overpic}[width=0.99\linewidth, trim={0.1cm, 0.25cm, 0.20cm, 0.2cm},clip]{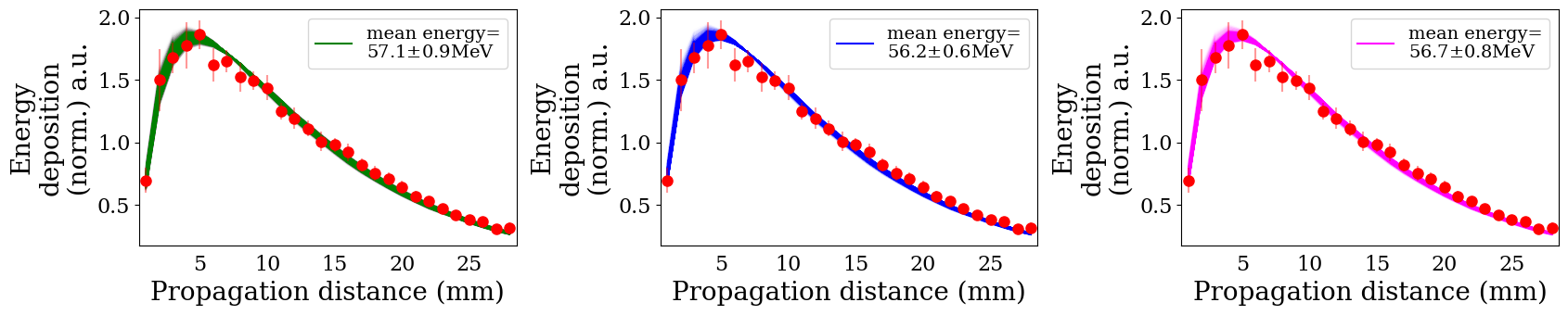}
\put(-2,18){b)}
\end{overpic}}
\end{subfigure}
\caption{\textbf{Bayesian inference results for shot 10.} Similar to figure~\ref{fig:post-coll_espec_no_chirp_1}. The distribution of $\tilde{a}_{0}$ inferred by the classical, quantum-continuous and quantum-stochastic models had mean and standard deviation, $\langle \tilde{a}_{0}\rangle=10.6\pm1.4$ and $\sigma_{a_0}=8.6\pm2.3$, $\langle \tilde{a}_{0}\rangle=14.5\pm2.2$ and $\sigma_{a_0}=7.7\pm1.7$ and $\langle \tilde{a}_{0}\rangle=13.3\pm1.9$ and $\sigma_{a_0}=8.6\pm2.3$, respectively. a) The measured post-collision electron spectrum, with $\langle E \rangle=\qty[parse-numbers = false]{617.3 \pm 0.0 (12.4)}{\mega\electronvolt}$ and $P_{70}=\qty[parse-numbers = false]{0.93 \pm 0.00 (0.01)}{\per\giga\electronvolt}$. The neural network predictions for the pre-collision electron spectrum have $\langle E \rangle=\qty[parse-numbers = false]{609.9 \pm 11.7 (12.1)}{\mega\electronvolt}$ and $P_{70}=\qty[parse-numbers = false]{1.3 \pm 0.19 (0.01)}{\per\giga\electronvolt}$. b) The measured photon spectrum, with mean energy \SI[separate-uncertainty]{68(7)}{\mega\electronvolt}.
\label{fig:post-coll_spec_no_chirp_3}}
\end{figure}

\begin{figure}[ht!]
\centering
\begin{subfigure}[t]{0.99\textwidth}
{\begin{overpic}[width=0.99\linewidth, trim={0.1cm, 0.20cm, 0.20cm, 0.22cm},clip]{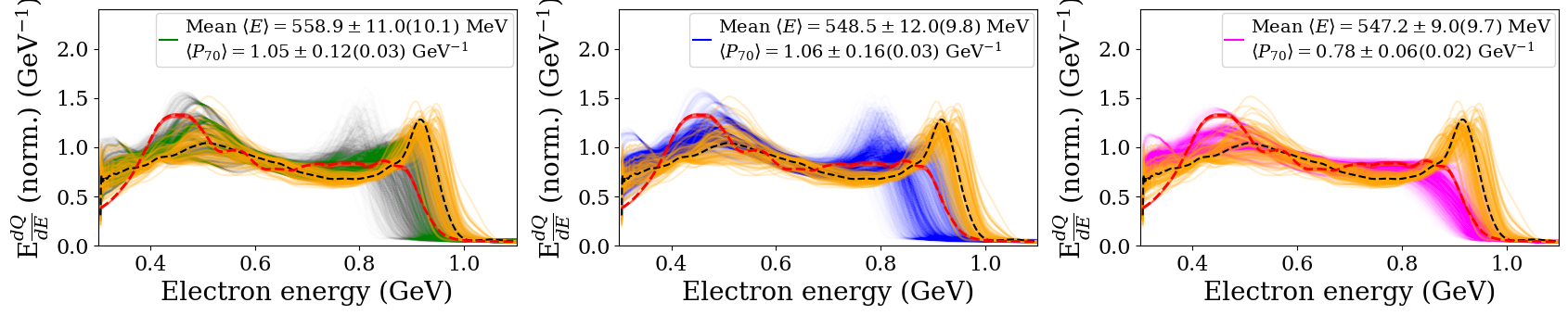}
\put(-2,18){a)}
\put(14,20.5){Classical}
    \put(40.5,20.5){Quantum-continuous}
	\put(74.5, 20.5){Quantum-stochastic}
\end{overpic}}
\end{subfigure}
\vspace{0.15em}
\\
 \begin{subfigure}[t]{0.99\textwidth}
{\begin{overpic}[width=0.99\linewidth, trim={0.1cm, 0.25cm, 0.20cm, 0.2cm},clip]{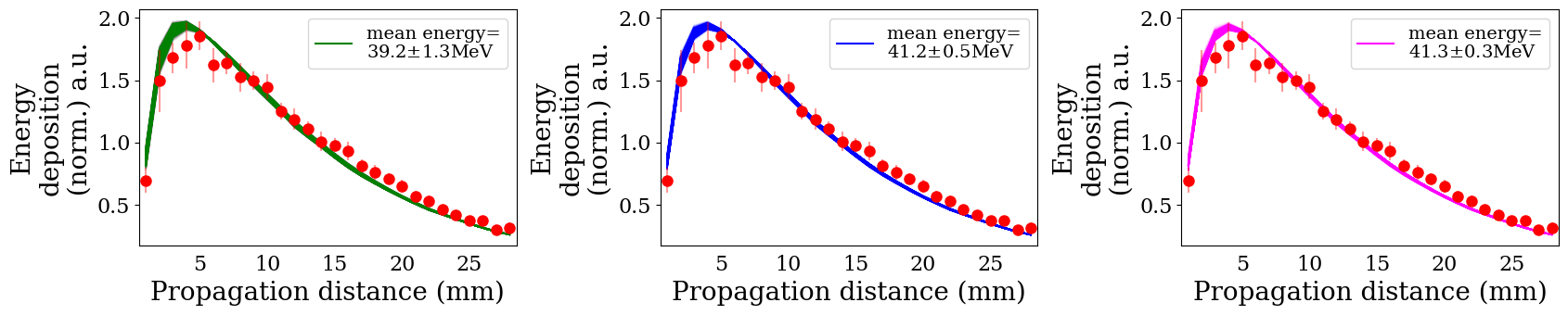}
\put(-2,18){b)}
\end{overpic}}
\end{subfigure}
\caption{\textbf{Bayesian inference results for shot 3.} Similar to figure~\ref{fig:post-coll_espec_no_chirp_1}. The distribution of $\tilde{a}_{0}$ inferred by the classical, quantum-continuous and quantum-stochastic models had mean and standard deviation, $\langle \tilde{a}_{0}\rangle=6.8\pm1.3$ and $\sigma_{a_0}=1.4\pm0.4$, $\langle \tilde{a}_{0}\rangle=7.6\pm1.0$ and $\sigma_{a_0}=0.5\pm0.1$ and $\langle \tilde{a}_{0}\rangle=7.5\pm0.7$ and $\sigma_{a_0}=0.5\pm0.1$, respectively. a) The measured post-collision electron spectrum, with $\langle E \rangle=\qty[parse-numbers = false]{564.9 \pm 0.0 (10.4)}{\mega\electronvolt}$ and $P_{70}=\qty[parse-numbers = false]{0.85 \pm 0.00 (0.01)}{\per\giga\electronvolt}$. The neural network predictions for the pre-collision electron spectrum have $\langle E \rangle=\qty[parse-numbers = false]{581.4 \pm 11.7 (11.0)}{\mega\electronvolt}$ and $P_{70}=\qty[parse-numbers = false]{1.23 \pm 0.08 (0.02)}{\per\giga\electronvolt}$. b) The measured photon spectrum, with mean energy \SI[separate-uncertainty]{63(14)}{\mega\electronvolt}.
\label{fig:post-coll_spec_no_chirp_2}}
\end{figure}

\begin{figure}[ht!]
\centering
\begin{subfigure}[t]{0.99\textwidth}
{\begin{overpic}[width=0.99\linewidth, trim={0.1cm, 0.20cm, 0.20cm, 0.22cm},clip]{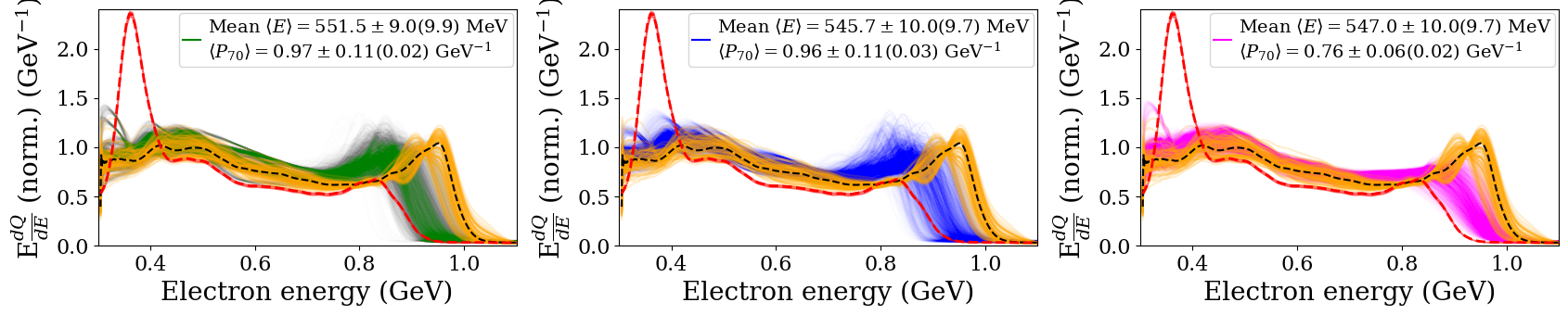}
\put(-2,18){a)}
\put(14,20.5){Classical}
    \put(40.5,20.5){Quantum-continuous}
	\put(74.5, 20.5){Quantum-stochastic}
\end{overpic}}
\end{subfigure}
\vspace{0.15em}
\\
 \begin{subfigure}[t]{0.99\textwidth}
{\begin{overpic}[width=0.99\linewidth, trim={0.1cm, 0.25cm, 0.20cm, 0.2cm},clip]{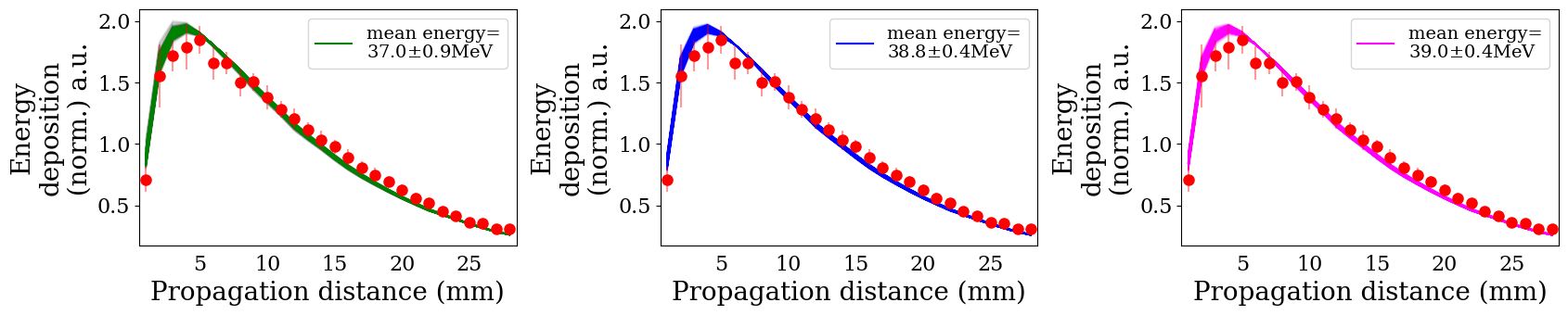}
\put(-2,18){b)}
\end{overpic}}
\end{subfigure}
\caption{\textbf{Bayesian inference results for shot 1.} Similar to figure~\ref{fig:post-coll_espec_no_chirp_1}. The distribution of $\tilde{a}_{0}$ inferred by the classical, quantum-continuous and quantum-stochastic models had mean and standard deviation, $\langle \tilde{a}_{0}\rangle=6.7\pm1.1$ and $\sigma_{\tilde{a}_{0}}=1.1\pm0.3$, $\langle \tilde{a}_{0}\rangle=8.1\pm1.2$ and $\sigma_{\tilde{a}_{0}}=1.1\pm0.3$, $\langle \tilde{a}_{0}\rangle=7.8\pm1.1$ and $\sigma_{\tilde{a}_{0}}=0.7\pm0.2$, respectively. a) The measured post-collision electron spectrum, with $\langle E \rangle=\qty[parse-numbers = false]{495.6 \pm 0.0 (8.0)}{\mega\electronvolt}$ and $P_{70}=\qty[parse-numbers = false]{0.67 \pm 0.00 (0.01)}{\per\giga\electronvolt}$. The neural network predictions for the pre-collision electron spectrum have $\langle E \rangle=\qty[parse-numbers = false]{572.1 \pm 6.9 (10.6)}{\mega\electronvolt}$ and $P_{70}=\qty[parse-numbers = false]{1.07 \pm 0.07 (0.02)}{\per\giga\electronvolt}$. b) The measured photon spectrum, with mean energy \SI[separate-uncertainty]{64(6)}{\mega\electronvolt}. 
\label{fig:post-coll_spec_no_chirp_8}}
\end{figure}

\begin{figure}[ht!]
\centering
\begin{subfigure}[t]{0.99\textwidth}
{\begin{overpic}[width=0.99\linewidth, trim={0.1cm, 0.20cm, 0.20cm, 0.22cm},clip]{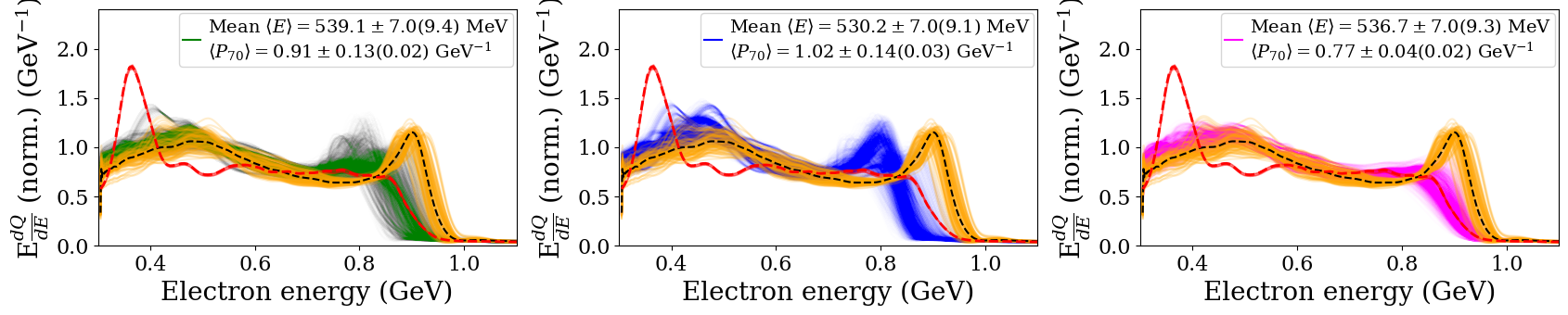}
\put(-2,18){a)}
\put(14,20.5){Classical}
    \put(40.5,20.5){Quantum-continuous}
	\put(74.5, 20.5){Quantum-stochastic}
\end{overpic}}
\end{subfigure}
\vspace{0.15em}
\\
 \begin{subfigure}[t]{0.99\textwidth}
{\begin{overpic}[width=0.99\linewidth, trim={0.1cm, 0.25cm, 0.20cm, 0.2cm},clip]{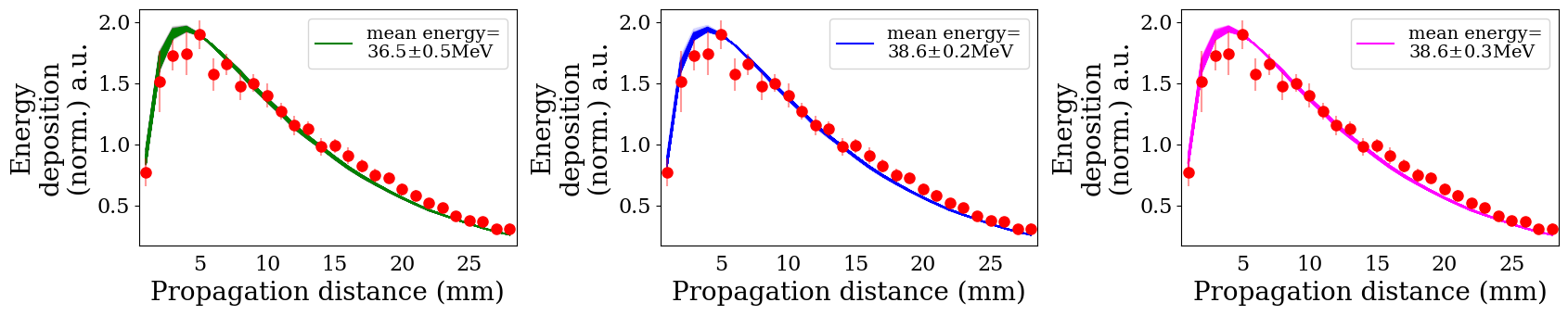}
\put(-2,18){b)}
\end{overpic}}
\end{subfigure}
\caption{\textbf{Bayesian inference results for shot 7.} Similar to figure~\ref{fig:post-coll_espec_no_chirp_1}. The distribution of $\tilde{a}_{0}$ inferred by the classical, quantum-continuous and quantum-stochastic models had mean and standard deviation, $\langle \tilde{a}_{0}\rangle=6.6\pm0.9$ and $\sigma_{a_0}=1.4\pm0.3$, $\langle \tilde{a}_{0}\rangle=7.2\pm0.5$ and $\sigma_{a_0}=0.4\pm0.1$ and $\langle \tilde{a}_{0}\rangle=7.2\pm$0.7 and $\sigma_{a_0}=0.5\pm0.1$, respectively. a) The measured post-collision electron spectrum, with $\langle E \rangle=\qty[parse-numbers = false]{524.9 \pm 0.0 (8.9)}{\mega\electronvolt}$ and $P_{70}=\qty[parse-numbers = false]{0.71 \pm 0.00 (0.01)}{\per\giga\electronvolt}$. The neural network predictions for the pre-collision electron spectrum have $\langle E \rangle=\qty[parse-numbers = false]{563.0 \pm 7.2 (10.3)}{\mega\electronvolt}$ and $P_{70}=\qty[parse-numbers = false]{1.15 \pm 0.04 (0.02)}{\per\giga\electronvolt}$. b) The measured photon spectrum, with mean energy \SI[separate-uncertainty]{77(30)}{\mega\electronvolt}.
\label{fig:post-coll_spec_no_chirp_9}}
\end{figure}

\clearpage

\appendix
\renewcommand{\thesection}{\Roman{section}}
\renewcommand{\thesubsection}{\Roman{subsection}.\arabic{subsection}}

\section{Extended Material}\label{sup:shot_selection_fig}

\subsection*{Frequentist analysis}\label{supp_freq_anal}


\begin{figure}[ht!]
\centering
{\begin{overpic}[width=1.0\textwidth, trim={0.2cm, 0.2cm, 0.2cm, 0.2cm},clip]{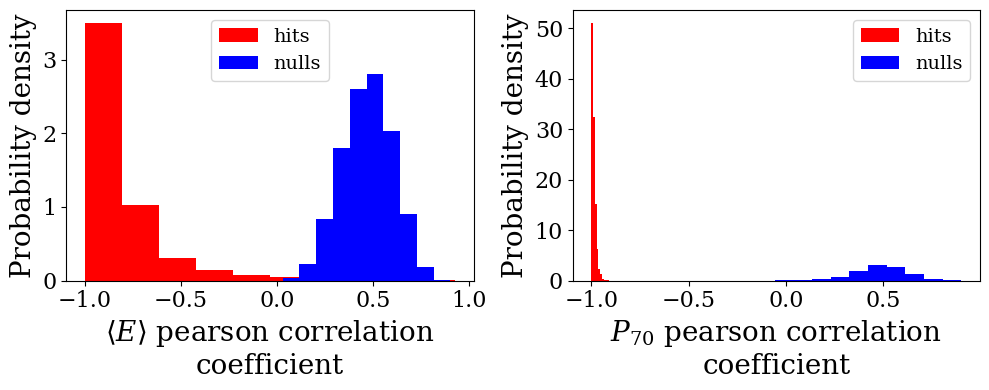}
  \put(-2,37){a)}
   \put(48,37){b)}
\end{overpic}}
\caption{Distributions of Pearson correlation coefficients calculated by boostrapping the binned hits and nulls shown in figure~\ref{fig:eloss_hists}.
\label{fig:corr_coeffs}}
\end{figure}

\begin{figure}[ht!]
\centering
\begin{subfigure}[t]{0.49\textwidth}
{\begin{overpic}[width=0.99\linewidth, trim={0.3cm, 0.20cm, 16.7cm, 0.22cm},clip]{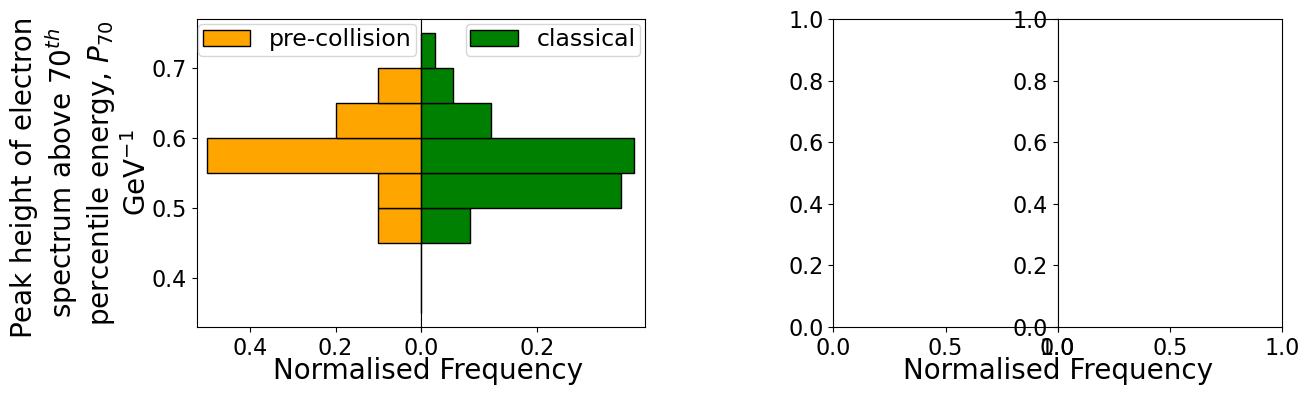}
\put(-2,60){a)}
\end{overpic}}
\end{subfigure}
\hfill
\begin{subfigure}[t]{0.49\textwidth}
{\begin{overpic}[width=0.99\linewidth, trim={0.25cm, 0.25cm, 16.7cm, 0.2cm},clip]{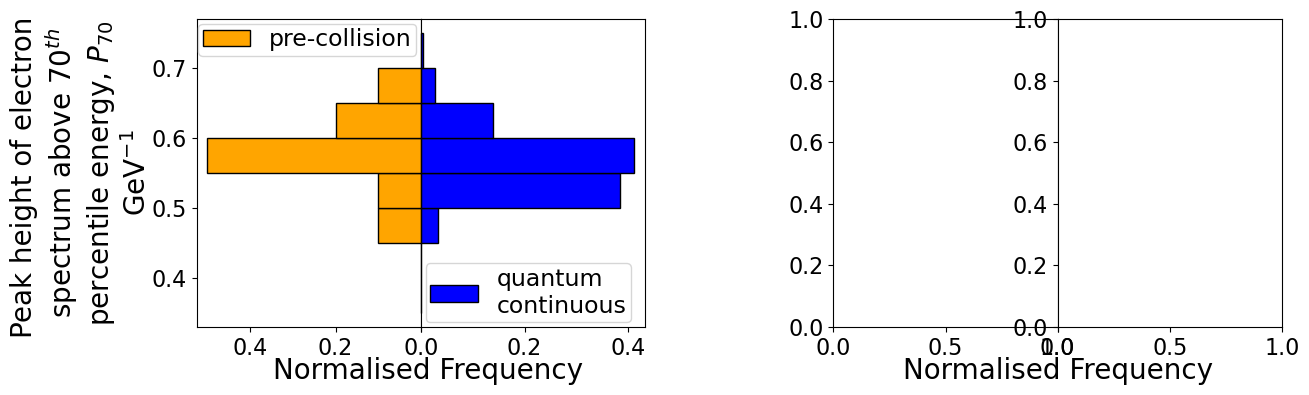}
\put(-2,60){b)}
\end{overpic}}
\end{subfigure}
\vspace{0.80em}
\\
 \begin{subfigure}[t]{0.49\textwidth}
{\begin{overpic}[width=0.99\linewidth, trim={0.25cm, 0.25cm, 16.7cm, 0.2cm},clip]{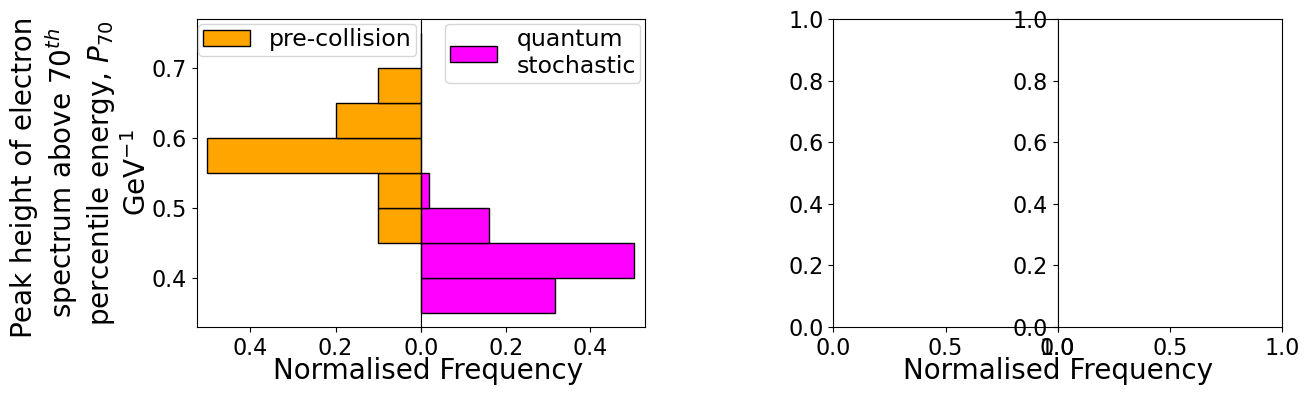}
\put(-2,60){c)}

\end{overpic}}
\end{subfigure}
\caption{\textbf{Effect of varying collision parameters on high-energy peak in electron spectrum ($P_{70}$)}.
Pre-collision and simulated post-collision distributions of $P_{70}$ predicted by the classical, quantum-continuous and quantum-stochastic models of radiation reaction for the ten pre-collision electron spectra predicted by the neural network. For each pre-collision spectrum, twenty-five collisions were simulated for varying $a_0$, $Z_d$ and $\tau_e$, drawn from the following gaussian distributions: $a_0 \sim \mathcal{N}(10, 2)$, $Z_d[\SI{}{\femto\second}] \sim \mathcal{N}(0, 30)$, $\tau_e[\SI{}{\femto\second}] \sim \mathcal{N}(40, 10)$. The means of the post-collision distributions of $P_{70}$ for the classical, quantum-continuous and quantum-stochastic models lie $8\sigma$, $8\sigma$ and $9\sigma$ below the pre-collision distribution mean, respectively, demonstrating that all radiation reaction models predict a reduction in $P_{70}$ when the electron beam interacts with different laser intensities during the collision.
\label{fig:peakiness_demo}}
\end{figure}


\clearpage
\bibliographystylesupp{unsrt}
\bibliographysupp{supp}

\end{document}